\documentclass[article,amsmath,amssymb,aps,floatfix,superscriptaddress]{revtex4-2}

\usepackage[margin=1in]{geometry}
\usepackage{tabularx} 
\usepackage{subfig}
\usepackage{amsmath}
\usepackage{amsfonts}
\usepackage{amsthm}
\usepackage{amssymb}
\usepackage{bbold}
\usepackage{graphicx}
\usepackage{booktabs}
\usepackage{dsfont}
\usepackage{calrsfs}
\usepackage{bbm}
\usepackage{enumerate}
\usepackage{listings}
\usepackage{mathrsfs,amsbsy}
\usepackage[english]{babel} 
\usepackage{setspace}
\usepackage{empheq}
\usepackage{hyperref}
\hypersetup{
	colorlinks=true,
	linkcolor=blue,
	filecolor=magenta,      
	urlcolor=cyan,
}

\allowdisplaybreaks

\newcommand{\beq}{\begin{equation}}
	\newcommand{\eeq}{\end{equation}}

\newcommand{\vt}[1]{\boldsymbol{#1}}
\newcommand{\overbar}[1]{\mkern 1.5mu\overline{\mkern-1.5mu#1\mkern-1.5mu}\mkern 1.5mu}
\DeclareMathAlphabet\mathbfcal{OMS}{cmsy}{b}{n}

\newcommand{\expect}[1]{\left\langle #1 \right\rangle}

\begin{document}

\title{Reconfigurable quantum phononic circuits via piezo-acoustomechanical interactions}

\author{Jeffrey C. Taylor}
\altaffiliation{These authors contributed equally to this work.}
\affiliation{Sandia National Laboratories, Albuquerque, New Mexico 87123, USA}

\author{Eric Chatterjee}
\altaffiliation{These authors contributed equally to this work.}
\affiliation{Sandia National Laboratories, Livermore, California 94550, USA}

\author{William F. Kindel}
\affiliation{Sandia National Laboratories, Albuquerque, New Mexico 87123, USA}

\author{Daniel Soh}
\affiliation{Sandia National Laboratories, Livermore, California 94550, USA}

\author{Matt Eichenfield}
\altaffiliation{Corresponding author: meichen@sandia.gov}
\affiliation{Sandia National Laboratories, Albuquerque, New Mexico 87123, USA}

\begin{abstract}

We show that piezoelectric strain actuation of acoustomechanical interactions can produce large phase velocity changes in an existing quantum phononic platform: aluminum nitride on suspended silicon. Using finite element analysis, we demonstrate a piezo-acoustomechanical phase shifter waveguide capable of producing $\pm\pi$ phase shifts for GHz frequency phonons in 10s of \textmu m with 10s of volts applied. Then, using the phase shifter as a building block, we demonstrate several phononic integrated circuit elements useful for quantum information processing. In particular, we show how to construct programmable multi-mode interferometers for linear phononic processing and a dynamically reconfigurable phononic memory that can switch between an ultra-long-lifetime state and a state strongly coupled to its bus waveguide. From the master equation for the full open quantum system of the reconfigurable phononic memory, we show that it is possible to perform “read” and “write” operations with over 90\% quantum state transfer fidelity for an exponentially decaying pulse.

\end{abstract}

\maketitle

\section{Introduction}

Phonons are becoming increasingly attractive for the processing of quantum information. Phononic quantum systems can be coupled to microwave frequency circuit QED systems via piezoelectric~\cite{arrangoiz-arriola_resolving_2019,arrangoiz-arriola_coupling_2018,jiang_efficient_2020,mirhosseini_quantum_2020,fink_efficient_2020,sletten_resolving_2019,oconnell_quantum_2010,chu_quantum_2017,moores_cavity_2018,ramp_wavelength_2020,wu_microwave--optical_2020,forsch_microwave--optics_2020,stannigel_optomechanical_2010,bochmann_nanomechanical_2013,vainsencher_bi-directional_2016,balram_coherent_2016,chu_quantum_2017} and electromechanical~\cite{palomaki_entangling_2013,palomaki_coherent_2013,mcgee_mechanical_2013,kerckhoff_tunable_2013,bagci_optical_2014,andrews_bidirectional_2014,suchoi_intermittency_2014,fink_quantum_2016,burns_bidirectional_2017,viennot_phonon-number-sensitive_2018,higginbotham_harnessing_2018,van_laer_electrical_2018,moaddel_haghighi_sensitivity-bandwidth_2018,mittal_reducing_2020,regal_sensing_2020} interactions, and yield long decoherence times when operated at cryogenic temperatures. For example, single crystal suspended silicon membranes cooled to cryogenic temperatures have been shown to have phononic lifetimes that are more than 4 orders of magnitude larger than those of superconducting circuit qubits at the same frequency~\cite{maccabe_nano-acoustic_2020,ren_two-dimensional_2020}. Moreover, when simultaneously localized with photons in optomechanical crystals~\cite{eichenfield_optomechanical_2009,chan_optimized_2012}, phonons interact strongly enough to have their quantum states teleported over optical fiber via optomechanical interactions~\cite{cohen_optical_2013,riedinger_remote_2018}, thus presenting a route to optical distribution and networking of microwave frequency quantum information~\cite{bochmann_nanomechanical_2013,vainsencher_bi-directional_2016,balram_coherent_2016,forsch_microwave--optics_2020,jiang_efficient_2020,mirhosseini_quantum_2020,neuman_phononic_2020,ramp_wavelength_2020,wu_microwave--optical_2020}. Since phonons can also interact strongly with electronic spins in solid state artificial atoms~\cite{jahnke_electronphonon_2015,lemonde_phonon_2018}, they may be capable of connecting many different quantum modalities over a quantum network~\cite{neuman_phononic_2020}.

However, the promise of phononics for quantum information applications hinges upon the enhanced functionality that could be achieved with active \cite{shao_electrical_2021}, as opposed to passive \cite{pechal_superconducting_2018, hann_hardware-efficient_2019, mirhosseini_quantum_2020}, phononic components. By dynamically reconfiguring phononic circuits, \textit{i.e.} through deterministic tuning of the speed of sound, new or enhanced functionalities arise that are not possible with static couplings and static phase velocities. For example, consider the case of coupling a superconducting qubit to a phononic cavity for storage. Transferring quantum information from the \textit{short} coherence time superconducting system to the intrinsically \textit{long} coherence time phononic cavity necessitates temporarily making the coupling rate to the phononic cavity larger than the decoherence rate of the qubit; after the quantum information has been fully coupled into the phononic state, the phononic cavity must be restored to its intrinsically long-lifetime state to allow for long duration storage with low decoherence. At present, the highest-$Q$ cavities suitable for such quantum information storage have been demonstrated in suspended Si phononic crystal membranes~\cite{maccabe_nano-acoustic_2020,ren_two-dimensional_2020}, and although electrical control of surface acoustic waves (SAW) has been demonstrated in LiNbO$_3$~\cite{shao_electrical_2021}, the loss channels affecting LiNbO$_3$ phononic crystal resonators at cryogenic temperature~\cite{wollack_loss_2021} have yet to be overcome to match the performance of Si resonators.

Here we consider how silicon, a non-piezoelectric material, can be mated with piezoelectric actuators to provide control over phonon phase velocity to yield useful devices for quantum information processing. This is analogous to the recent developments in piezoelectrically strain-tuned photonic integrated circuits~\cite{stanfield_cmos-compatible_2019,dong_high-speed_2021,tian_hybrid_2020,jin_piezoelectrically_2018}. Instead of employing the electro-acoustic effect of piezoelectric materials~\cite{shao_electrical_2021}, we show that acoustomechanical interactions—the interactions of phonons with the materials that guide them as those materials deform and strain—provide the necessary tunability and reconfigurability to enable such phononic circuits. Two principal acoustomechanical effects afford control of the phase velocity of phonons: a moving boundary effect and an acoustoelastic effect. These effects are analagous to the optomechanical moving boundary~\cite{balram_moving_2014, eichenfield_optomechanical_2009,eichenfield_modeling_2009, johnson_perturbation_2002,hossein-zadeh_observation_2007} and photoelastic~\cite{rakich_tailoring_2010,otterstrom_optomechanical_2018,gyger_observation_2020} effects, which can be used to control the phase velocity of photons. We employ an exemplary architecture, Sc$_x$Al$_{1-x}$N piezoelectric actuators integrated with silicon phononic crystal (PnC) circuits, to study how piezoelectrically-actuated strain can be used to produce these acoustomechanical effects on demand. There are already existing systems in which low-loss phonons are coupled to piezoelectric actuators~\cite{mirhosseini_quantum_2020,wu_microwave--optical_2020,arrangoiz-arriola_coupling_2018,arrangoiz-arriola_resolving_2019}. We provide a complete design and finite element method (FEM) analysis of an achievable phase shifting device architecture that employs a materials system that has been effectively demonstrated in prior research~\cite{mirhosseini_quantum_2020,ward_all-optical_2017,ghatge_high_2018,kittlaus_electrically_2021,pan_thin-film_2010}. We use the phase shifting waveguide as a building block to describe several phononic integrated circuit components useful for quantum information processing, including programmable multi-mode interferometers (PMMIs), reconfigurable quantum memory qubits, and addressable quantum memory registers. For the quantum memory, we use the so-called Scattering-Lindblad-Hamiltonian (SLH) quantum network formalism~\cite{combes_slh_2017} to describe quantum information transfer between a flying phononic qubit in a waveguide and a tunable ultra-high-$Q$ phononic cavity. The SLH formalism yields a master equation for the full open quantum system that we use to optimize the classical control fields and achieve 90\% fidelity in the quantum information transfer.

\section{Principles of an acoustomechanical phase shifter}

To develop a strain-actuated phononic phase shifting device, we must account for how mechanical deformations impact the phase accumulated when a signal travels in a waveguide. In general, the phase $\phi$ accumulated by a phonon propagating a distance $L$ is given by the product of $L$ and the propagation constant $k$. By straining the waveguide, both $k$ and $L$ are altered, and the first-order difference in the acquired phase is
\begin{equation} \label{Eqn:deltaphi}
    \Delta\phi = \Delta k \cdot L + k \cdot \Delta L.
\end{equation}

\begin{figure}
	\centering
	\includegraphics[width=\textwidth]{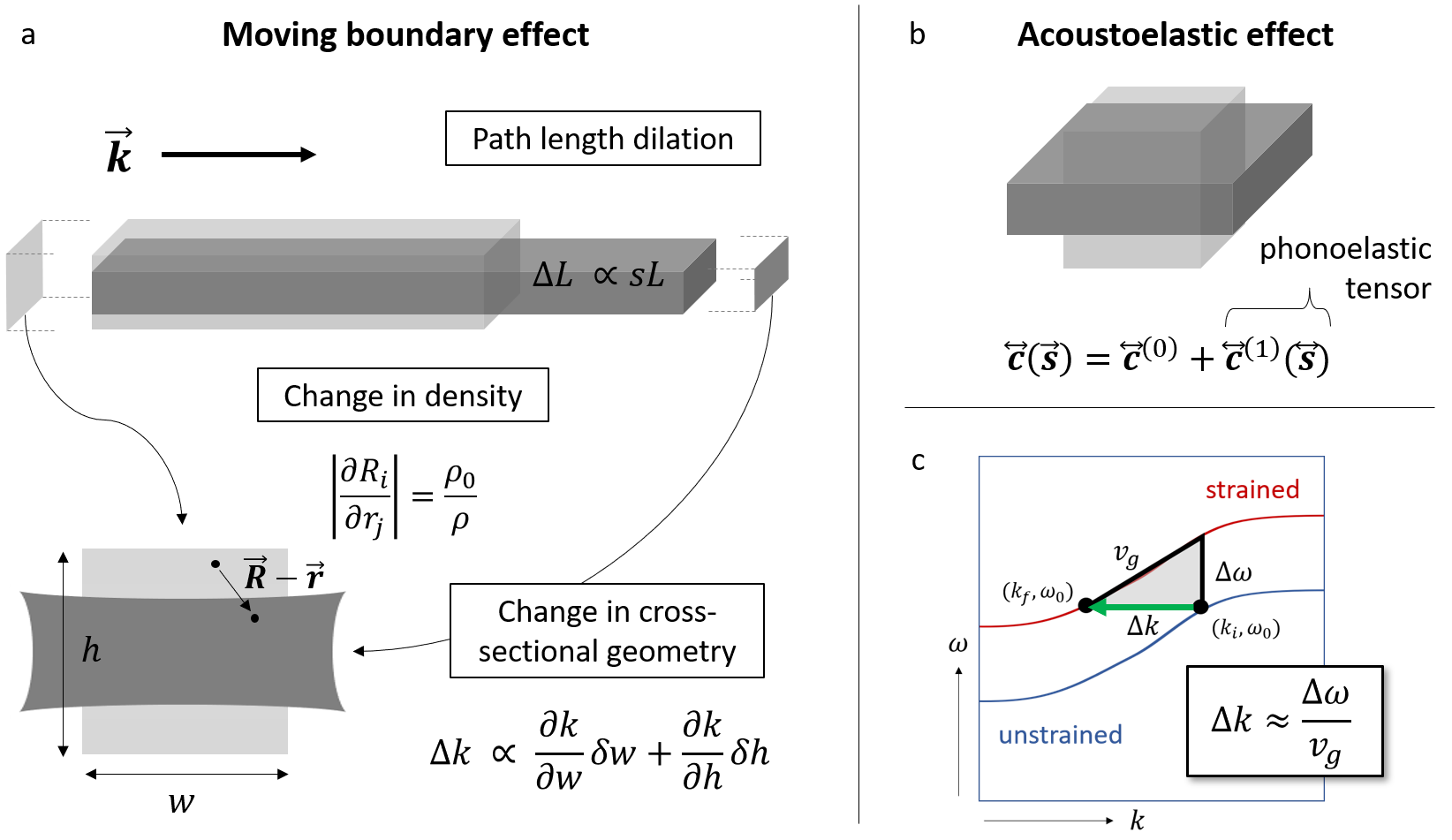}
	\caption{Acoustomechanical impact of strain on a waveguide. (a) The moving boundary effect describes the change in propagation constant with a modified cross-sectional geometry and density. The Poisson effect leads to an additional phase shift accumulated from the change in path length. (b) Single-crystal silicon behaves acoustoelastically at high strain, resulting in a nonlinear elasticity tensor. (c) Schematic of band perturbation under an applied strain. When the operating point shifts from ($k_i,\omega_0$) to ($k_f,\omega_0$) a phase shift will accrue over the length of the strained waveguide.} \label{fig:acoustomechanical_effects}
\end{figure}

\figureautorefname{ \Ref{fig:acoustomechanical_effects}} outlines how the acoustomechanical effects contribute to the perturbations $\Delta k$ and $\Delta L$. The moving boundary effect [\figureautorefname{ \ref{fig:acoustomechanical_effects}}(a)] arises because the modal dispersion is a function of waveguide geometry. In a continuously periodic waveguide, the moving boundary effect is captured by the cross-sectional deformation and the change in path length, but in a discretely periodic structure the changing length of the unit cell is also pertinent. The moving boundary effect is intrinsically coupled to the change of density within the body, and both are captured by the Jacobian $J$ of the deformation, which is equal to the determinant of the deformation gradient tensor $\mathbf{A}$ and the inverse relative change in density: $J = \det(\mathbf{A}) = | \partial R_i/\partial r_j | = \rho_0/\rho$, where $\vec{R}$ and $\vec{r}$ represent the coordinates of a material particle in the strained and unstrained states respectively~\cite{thurston_third-order_1964}. The acoustoelastic effect [\figureautorefname{ \ref{fig:acoustomechanical_effects}}(b)] results from strain dependence in the elasticity tensor. Its scale is determined by intrinsic material properties which are captured by extending the constitutive relation for the elastic strain energy density $W$ to third order. Details regarding this hyperelastic constitutive model are included in the Supplemental Information (SI).  Acoustomechanical effects are analogous to optomechanical interactions in photonics~\cite{aspelmeyer_cavity_2014}, which also have both a material and geometric component: the photoelastic and optical moving boundary effects, respectively. 


\figureautorefname{ \Ref{fig:acoustomechanical_effects}}(c) outlines the influence of a static perturbation on the band structure of a phononic waveguide such that a point identified with an operating frequency $(k_i,\omega_0)$ becomes $(k_i,\omega_0+\Delta\omega)$. The propagation constant for a wave traveling in the perturbed structure at $\omega_0$ must change by $|\Delta k| = |k_f-k_i|= (d\omega/dk)^{-1}\Delta\omega=\Delta\omega/v_g$, where $v_g = (d\omega/dk)$ is the local group velocity. Therefore, the bandstructure\textemdash and in particular the group velocity\textemdash of the guided mode mediates the responsiveness of the waveguide to the acoustomechanical effects, a fact that we will exploit here.

In the finite element method (FEM) modeling that follows, we calculate $\Delta k$ by considering a unit cell of a discretely periodic waveguide being quasi-statically strained. With Floquet boundary conditions, we can use FEM to solve for the phonon dispersion as a function of $k$ across the first Brillouin zone and provide an estimate of $v_g$ as a function of $k$. Placing bias on the piezoelectric actuators in our simulation leads to a solution for $\Delta \omega$ relative to the unstrained state of the system. Floquet periodicity isolates the moving boundary effect within a single waveguide period from the overall change in path length since it provides a solution for the dispersion when the unit cell deforms both longitudinally and cross-sectionally in response to strain. We will show that the overall change in path length length $\Delta L$ is small and, at least in the specific example we model, leads to an offset in $\Delta \phi$ that is on the order of $1\%$ relative to the change in phase associated with $\Delta k$. Since $\Delta L$ depends on boundary conditions, the geometry of the piezoelectric loading, and the overall length of the waveguide, it must be estimated by specifying a complete device geometry and finding the stationary state of the system when bias is applied to the piezoelectric actuators.

\section{Piezoelectrically-actuated phononic phase shifter} \label{Sec:II}

To realize a phononic phase shifting device, we must specify a system geometry and assess the voltage dependence of the phase shift that is induced for signals in a waveguide. Two steps, corresponding to the two terms in \equationautorefname{ \Ref{Eqn:deltaphi}}, are required to estimate the effectiveness that our phononic waveguide system provides for inducing a phase shift. First, we consider a single period computational structure with infinitely periodic Floquet boundary conditions to estimate the phase shift induced by the acoustomechanical effects. This periodic system yields an estimate of $\Delta k \cdot L$. Then we construct a finite structure and solve for its stationary state when bias is applied to the piezoelectric actuators so that we can obtain a measure of $k \cdot \Delta L$. Since both terms in \equationautorefname{ \Ref{Eqn:deltaphi}} depend on the operating point in the guided band, the capacity for shifting the phase depends on this choice of $(k_0,\omega_0)$.

\begin{figure}
	\centering
	\includegraphics[width=\textwidth]{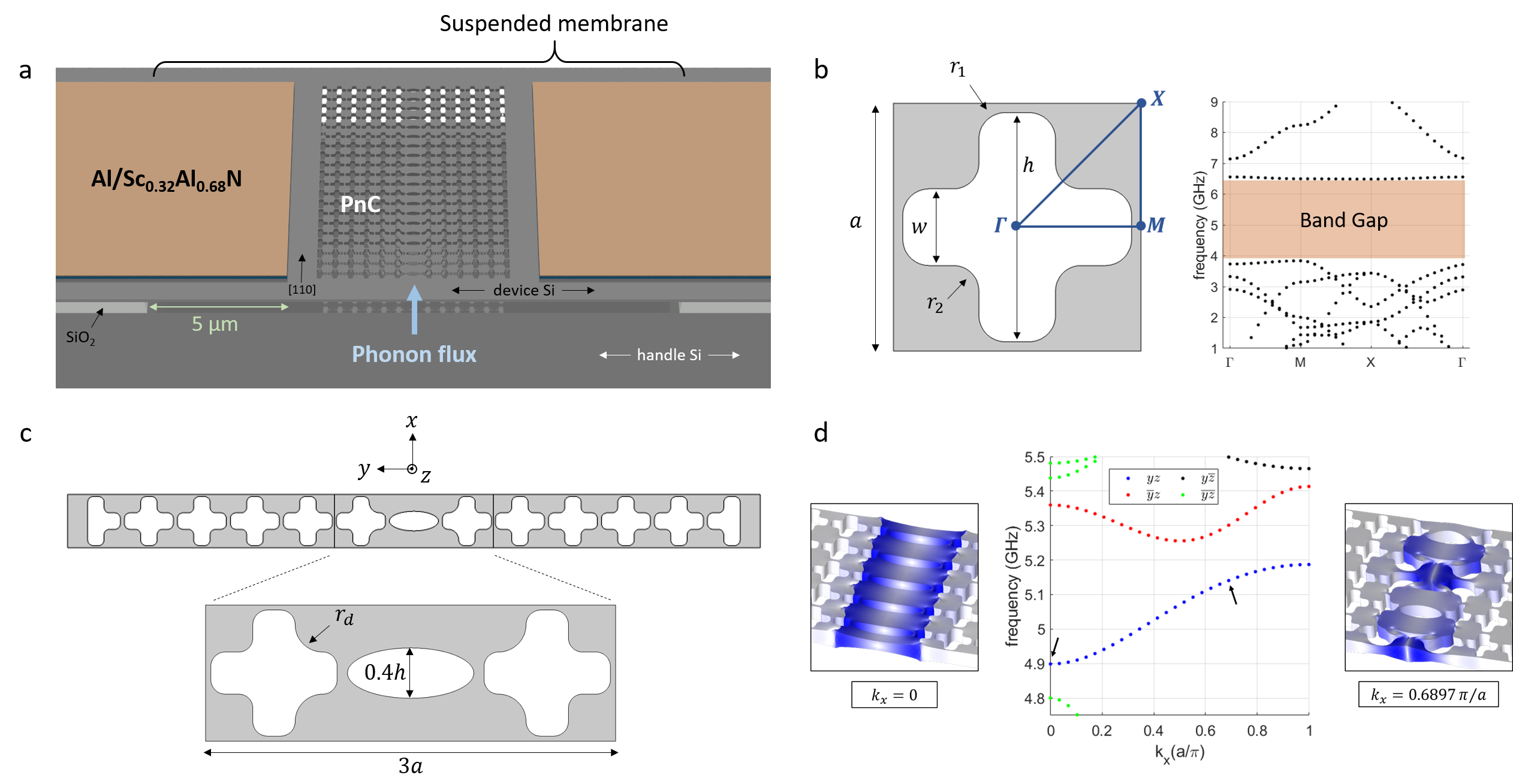}
	\caption{(a) Three-dimensional rendered view of a phase shifting waveguide that is 20 periods long. The waveguide and 5 \textmu m of actuator adjacent to it on either side are suspended. (b)  Geometrical definition of the plan-view structure of the PnC alongside the computed bandstructure. (c) Geometrical parameters for the ‘defect’ region. (d) Bandstructure for the waveguide showing that there is a fully symmetric mode (blue points) that is isolated in frequency and which is monotonically increasing as a function of $k_x$. The bands are colored according to their symmetry properties. The overbars in the legend denote inversion symmetry. The insets show the displacement profiles of the mode at the center of the Brillouin zone and at $k_x=0.6897\pi/a$.} \label{fig:phase_shifter_structure}
\end{figure}

\figureautorefname{ \Ref{fig:phase_shifter_structure}} presents our exemplary phase shifter system geometry. As in Ref.~\cite{mirhosseini_quantum_2020}, this structure could be fabricated on SOI wafers such that the functional portion of the device is situated on a suspended Si membrane where the intermediate oxide layer has been etched away, for example in a vapor HF process. We assume that the top device layer of the SOI is 250 nm thick. To ensure ultra-low propagation losses~\cite{maccabe_nano-acoustic_2020,alegre_quasi-two-dimensional_2011}, five periods of phononic crystal (PnC) cladding surround the waveguide defect, which spans 3 periods. The PnC structure is modeled after the acoustic shielding used in Ref.~\cite{maccabe_nano-acoustic_2020}, except for the difference in the thickness of the SOI wafer's device layer. The PnC lattice has a pitch $a=530$~nm and is defined by holes with geometrical parameters (expressed in units of $a$) of width $w=0.31$, height $h=0.925$, inner radius of curvature $r_1=0.1067$, and outer radius of curvature $r_2=0.1022$. The infinitely periodic two-dimensional structure yields a bandgap that extends for more than 1 GHz both below and above our targeted operating regime near 5 GHz.
The waveguide has a central elliptical hole with a major axis of size $h$ and a minor axis of 0.4$h$, and the PnC holes adjacent to the ellipse have outer radii of curvature increased to a maximal value of $r_d=(h-w)/2 - r_1$. The resulting guided band is isolated in frequency at approximately 5 GHz, has a symmetric displacement profile with respect to its $y$ and $z$ mirror planes, and has normal dispersion across the entire Brillouin zone. Sc$_{0.32}$Al$_{0.68}$N piezoelectric actuators are made adjacent to the waveguide to provide strain tuning, and the actuators are assumed to be suspended for 5 \textmu m on either side of the PnC. A spacing of 1 \textmu m exists between the PnC and the actuators. The actuators operate under the influence of a vertical field and require electrodes above and below the piezoelectric material, which is 250 nm thick. The bottom electrode of the actuators is assumed to be constructed in the Si device layer using high doping that yields metallic electronic conduction~\cite{pan_thin-film_2010} while the top electrode is constructed from 75 nm of Al.

\begin{figure}
	\centering
	\includegraphics[width=\textwidth]{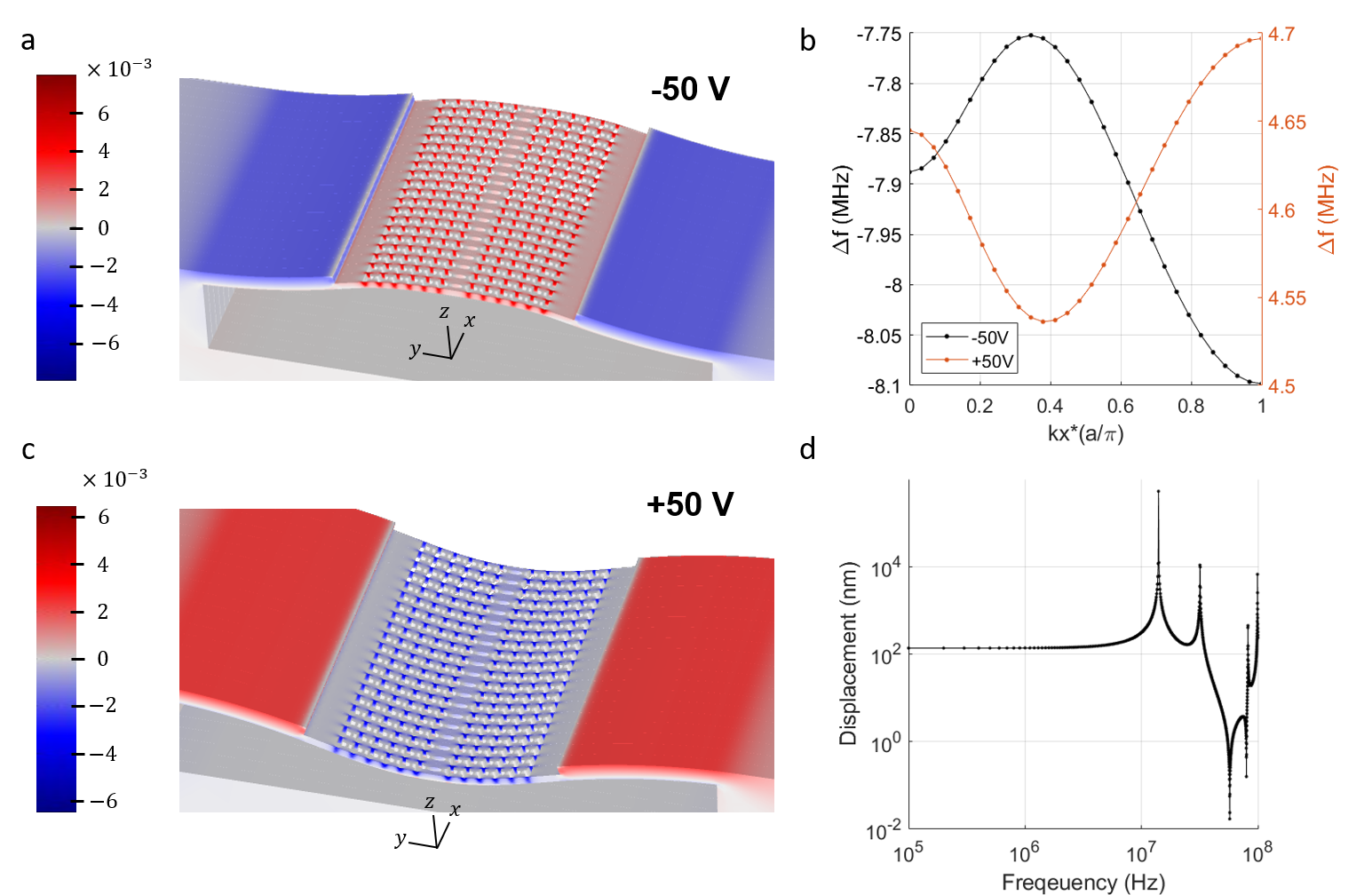}
	\caption{Simulations showing the effect of bias on the phase shifting waveguide structure described in \figureautorefname{ \Ref{fig:phase_shifter_structure}}, where (a,c) show the displacement at $-50$ V and $+50$ V with a 10x enhancement. The transverse component of strain field $s_{yy}$ is plotted according to the color scale. (b) Frequency shift $\Delta f$ evaluated across the entire Brillouin zone at both $-50$ V and $+50$ V. (d) The frequency response of the system where the total displacement for a representative point in the center of the waveguide is plotted versus frequency.}
	\label{fig:piezo_bias_effect}
\end{figure}

\figureautorefname{ \Ref{fig:piezo_bias_effect}} shows the strain induced by biases of $-50$ V and $+50$ V across the piezoelectric actuator. The mechanical responses of the system are shown in \figureautorefname{ \Ref{fig:piezo_bias_effect}}(a,c), with a 10x augmentation in the displacement and the transverse strain field $s_{yy}$. Bowing in the membrane arises due to the asymmetric loading caused by the top-only placement of the piezoelectric film, which is consistent with the limitations of multilayer microfabrication techniques. In \figureautorefname{ \Ref{fig:piezo_bias_effect}}(b), the variation in the frequency shift $\Delta f = \Delta\omega/2\pi$ across the Brillouin zone is plotted for both the $-50$ and $+50$ V biases, showing that the band frequencies shift by over a MHz at these biases. In addition, the 50 V transfer function for the displacement of a representative point in the center of the waveguide is shown in \figureautorefname{ \Ref{fig:piezo_bias_effect}}(d). The first resonance is found at 14 MHz, yielding a minimum switching time of 71 ns. 

\begin{figure}
    \centering
    \includegraphics[width=\textwidth]{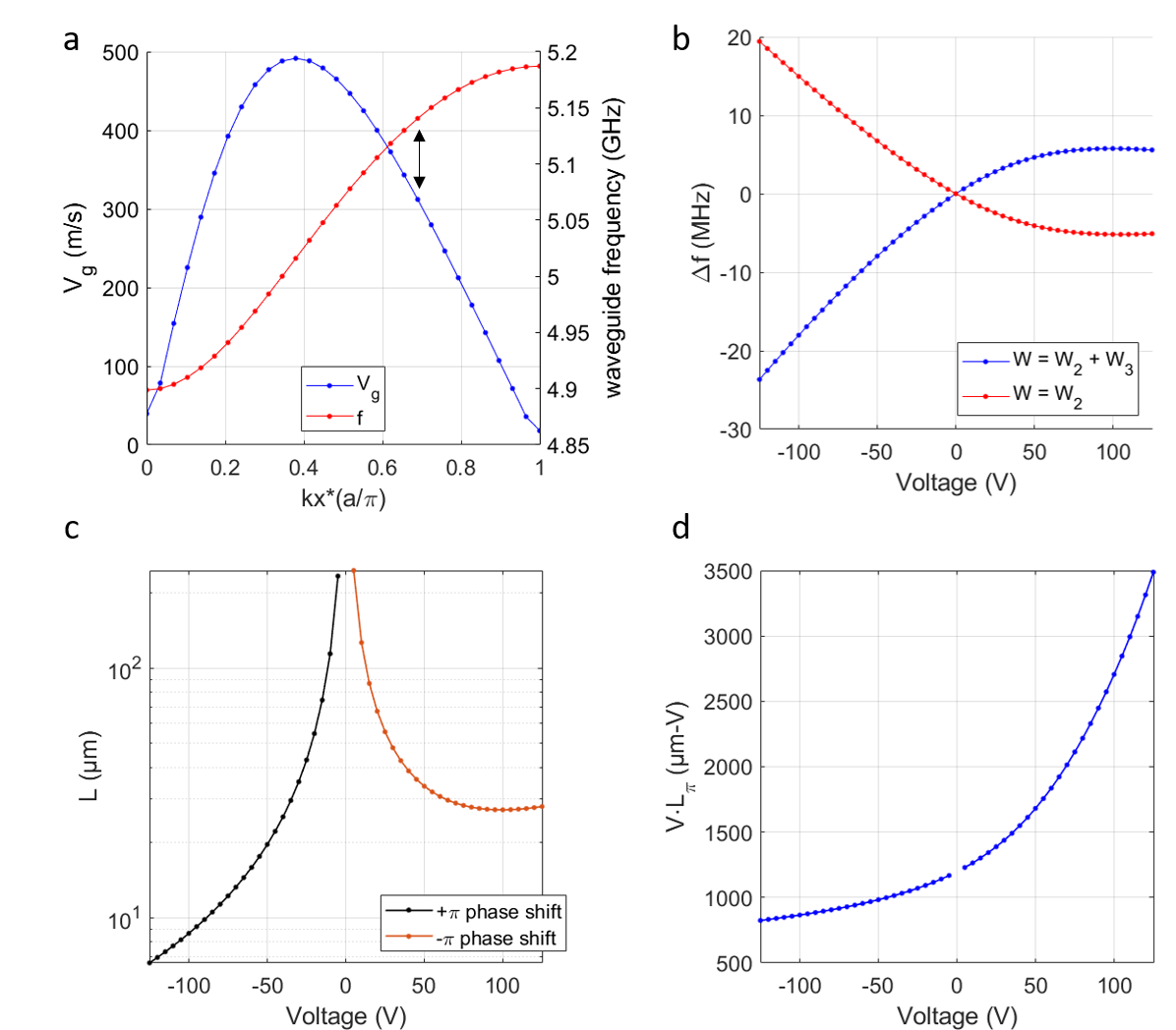}
    \caption{(a) The group velocity and frequency of the waveguide band, with an arrow designating the operating point at $k_x=0.6897 \pi/a$. (b) Voltage dependence of $\Delta f$ computed using two different constitutive relations. (c) Waveguide length $L_{\pm \pi}$ versus applied bias, showing the conditions required to achieve a $\pm \pi$ phase shift. (d) The figure of merit $(V\cdot L)_\pi$ plotted versus $V$, emphasizing the effect of the bowing deformation on the phase shift. A strain applied exclusively in the $y$-axis would yield a linear relation.}
    \label{fig:ps_op_pt}
\end{figure}

To analyze the acoustomechanical phase shift $\Delta k \cdot L$ we focus on a particular operating point at 5.1406 GHz with $k_x=0.6897 \pi/a$ and $v_g=312$ m/s. This operating point is designated with a double arrow in \figureautorefname{ \Ref{fig:ps_op_pt}}(a), which shows the group velocity $v_g$ plotted alongside a magnified plot of the guided modal frequencies. Biasing the actuators will cause a shift in frequency from which we can calculate $\Delta k$, as described in the previous section. To incorporate the acoustoelastic effect, we employ a hyperelastic material model with a constitutive relation for the elastic strain energy density $W$ that includes the effect of the third order elastic moduli~\cite{brugger_thermodynamic_1964}. In a hyperelastic material, $W$ exists such that $T_{ij}\equiv \partial W / \partial s_{ij}$~\cite{ogden_non-linear_1997}. When only the second order elastic strain energy is included and $W=W_2$, the hyperelastic constitutive relation is equivalent to the conventional definition of the elasticity tensor via the stress-strain relation $T_{ij}=c_{ijkl}s_{kl}$. Including the third order term $W_3$ in the elastic strain energy models the acoustoelastic effect and yields a strain-dependent correction to the stress tensor which we call the `phonoelastic' tensor, the components of which are derived in the SI for the case of diamond cubic Si. The variation of $\Delta f$ with $V$, both including ($W=W_2+W_3$) and excluding ($W=W_2$) the third order strain energy term is shown in \figureautorefname{ \Ref{fig:ps_op_pt}}(b). The acoustoelastic effect reverses the polarity and alters the magnitude of $\Delta f$, showing that optimization of phase shifts in these systems requires a careful design that takes into account both the acoustoelastic and moving boundary effects. A similar calculation for an alternative waveguide geometry with $a=2.07$ \textmu m, and which operates near 1.5 GHz, is presented in the SI and shows only a marginal contribution from $W_3$. In both cases, there is a significant nonlinearity in $\Delta f$ that arises due to the broken symmetry in the placement of the piezoelectric actuators. If the piezo load could be exclusively applied in the plane of the PnC membrane, the variation of $\Delta f$ with $V$ would be linear, which is also demonstrated in the SI.

The total phase shift accumulated over a length $L$ of strained waveguide is given by the product of $L\cdot \Delta k$. If we assume that the desired phase shift ranges from $-\pi$ to $+\pi$, we can express the device length needed as a function of the actuator bias,
\begin{equation}
	L(k,V)=\frac{v_g(k)}{2 \cdot |\Delta f(k,V)|}
\end{equation}
where the sign of $\Delta f$ determines whether the phase velocity increases or decreases. For the device described in \figureautorefname{s \Ref{fig:phase_shifter_structure} and \Ref{fig:piezo_bias_effect}}, a negative (positive) bias with tensile (compressive) $s_{yy}$ strain yields a forward (backward) phase shift because the frequency shift is negative (positive). Since the frequency shift is nonlinear, voltages of differing magnitudes are required for positive and negative phase shifts if the length is held constant, as illustrated in \figureautorefname{ \Ref{fig:ps_op_pt}}(c). In a linear system in which the piezo load could be exclusively applied in the plane of the PnC membrane, the figure of merit for a phase shifter would be the product $(V\cdot L)_\pi$ but this quantity does not yield a linear relation due to the underlying nonlinearity in $\Delta f$, as shown in \figureautorefname{ \Ref{fig:ps_op_pt}}(d). Despite this bowing induced nonlinearity, the 5 GHz system described here is stiff enough to achieve compressive stress at positive bias, although the phase shifting response in compression is muted. As the extent of the piezoelectric transducer's overhang increases, the tensile phase shift at negative bias improves at the expense of the compressive phase shift at positive bias and the switching time. These effects are demonstrated by the 1.5 GHz system described in the SI, which we present with a membrane overhang that extends for more than 5x longer than we have specified in the 5 GHz system. Compared with the SAW devices presented in Ref.~\cite{shao_electrical_2021}, we neglect the propagation loss from the figure of merit due to the high $Q$ expected in a suspended single-crystal Si PnC resonator, but we acknowledge that in future experimental demonstrations fabrication defects will inevitably impact our low $v_g$ waveguides~\cite{hughes_extrinsic_2005}.

To assess the purely mechanical phase shift arising from the change in path length, we construct a finite structure in which the piezoelectric actuator extends for the central 20-periods of a waveguide that is 220 periods long. The details of this calculation are provided in the SI. For the operating point at $k=0.6897\pi/a$ and a voltage of $-50$ V, this finite structure yields a phase shift due to the mechanical change in path length $k \cdot \Delta L$ that is approximately $0.1^\circ$. Since $\Delta k \cdot L\approx 97^\circ$ over 20 periods of the periodic phase shifter structure at $-50$ V, we conclude that the phase shift captured by the periodic computational system will be dominant in our phase-shifting device structure.

\section{Quantum information processing devices enabled by the phononic phase shifter}

In this section we describe how our phononic phase shifting device platform can be applied to store, route and process quantum information. Storage is accomplished by implementing a phononic cavity with tunable coupling. Information can be routed and processed with Mach-Zehnder interferometers. In addition, our waveguide geometry lends itself to fine tuning of dispersion, and suggests the construction of mirrors with tunable reflectivity (described in the SI). When these elemental phononic devices are coupled to superconducting circuits via piezoelectric interactions~\cite{mirhosseini_quantum_2020}, more advanced quantum information processing operations result. For example, we later describe how an addressable quantum memory arises from using Mach-Zehnder interferometers as unitary transformation operators on waveguide spatial modes. The interferometers can direct signals into qubit memories comprised of cavities with tunable coupling. In addition, since the SOI fabrication process is scalable, our device platform could be used to yield large networks of Mach-Zehnder interferometers for universal linear phononics, as has been shown with photonics~\cite{carolan_universal_2015}.

\subsection{Phononic cavity with tunable coupling}

\begin{figure}[ht]
    \centering
    \includegraphics[width=\textwidth]{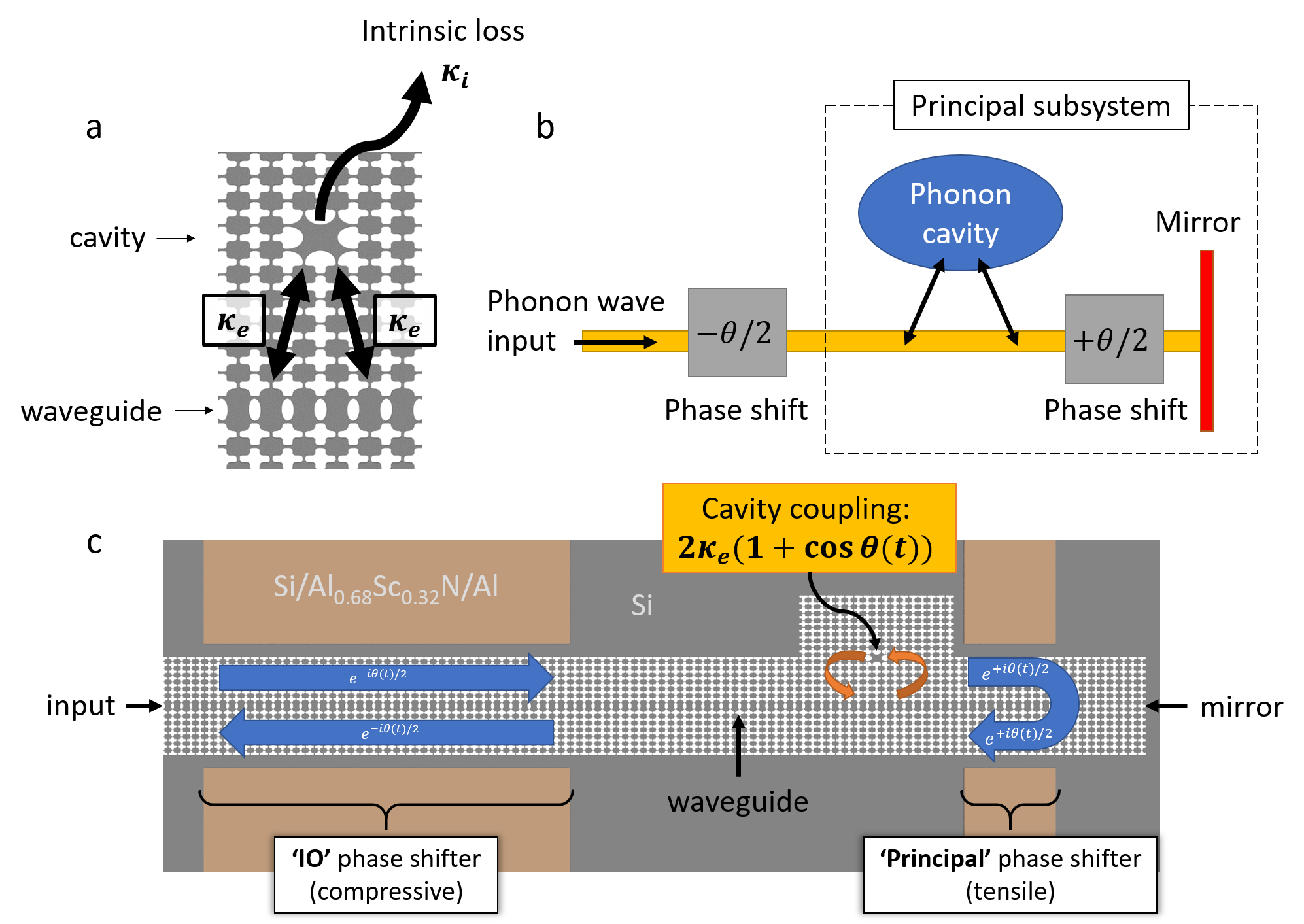}
    \caption{(a) Phononic cavity with an isolated resonance at 5.14 GHz. The cavity has an intrinsic loss rate $\kappa_i \sim 1~$Hz and an extrinsic loss rate $\kappa_e$ which can be tuned by adjusting the spacing between the cavity and the waveguide. (b) Block diagram showing the essential components of the system. (c) Schematic structure of a phononic quantum memory device.}
    \label{fig:Phononic_memory_architecture}
\end{figure}

To construct a phononic cavity with tunable coupling suitable for a quantum memory device, we combine a pair of the phononic phase shifters described above with a cavity of high intrinsic $Q$, following an analogous photonic device described in Ref.~\cite{tanaka_dynamic_2007}. In this scheme, a high-$Q$ cavity serves as a storage medium where the cavity's coupling to a bus waveguide can be controlled via interferometrically combining the leakage channels in two different directions. Consider the cavity evanescently coupled to a phononic waveguide shown in \figureautorefname{ \Ref{fig:Phononic_memory_architecture}}(a). This cavity has an isolated resonance at 5.14 GHz with a fully symmetric displacement profile. Thus it can couple with the operating point in the guided band of our phase shifting waveguide as described in \sectionautorefname{ \ref{Sec:II}}. Based on recent results~\cite{maccabe_nano-acoustic_2020,ren_two-dimensional_2020}, we assume that the cavity is at cryogenic temperatures and has intrinsic $Q$ in excess of $10^{10}$ so that its intrinsic loss rate $\kappa_i$ is on the order of $2\pi \times 1$ Hz. The evanescent coupling is characterized by a raw output coupling rate $\kappa_e$ in each direction along the waveguide. To capture a flying phonon incident on the cavity from the waveguide, $\kappa_e$ must be greater than bandwidth of that flying phonon. This can be achieved by adjusting the proximity of the cavity and waveguide through the phononic lattice or by modifying the geometry of the lattice between them. 

A block diagram of the essential components for the memory architecture is shown in \figureautorefname{ \Ref{fig:Phononic_memory_architecture}}(b) and a schematic of a possible realization is shown in \figureautorefname{ \Ref{fig:Phononic_memory_architecture}}(c). There are two phase shifters involved, but the essential operating components for the memory are constituted by the subsystem that is demarcated with a dashed line in \figureautorefname{ \Ref{fig:Phononic_memory_architecture}}(b). Along with the mirror, this subsystem consists of the cavity and the `principal' phase shifter to the right of the cavity. Quantum information traveling along the waveguide interacts with the phonon cavity twice, once traveling to the right and a second time traveling from the left after reflecting from the mirror. The system dynamics result from an interference of 3 waves at the left output port of the cavity: (1) the input wave that is transmitted past the cavity, (2) the raw rightward-traveling output from the phonon cavity, and (3) the raw leftward-traveling output. Waves (1) and (2) have a round trip propagation phase that is dependent on the state of the principal phase shifter, allowing for the interference between each of these waves and wave (3) to be controlled. The system functions as a memory when the phase shift induced by the principal phase shifter during the round trip is tuned so that the net leftward-traveling output from the cavity is nulled and the net input coupling is maximized. The larger, `IO' phase shifter on the left side of the schematic functions in opposition to the principal phase shifter and ensures that the output from the overall system is in phase with the input.

To define a nominal value of $\kappa_e$, we consider coupling our memory to a superconducting transmon qubit, e.g. via the piezoelectric effect~\cite{mirhosseini_quantum_2020}. For simplicity, we assume that the input signal from the qubit arises from a mode with a fixed coupling and total attenuation rate $r$ so that the input signal has an exponentially decaying profile with power decay rate $r$. To preserve the coherence between the input signal and qubit system, this decay must happen faster than the qubit's coherence time $\mathrm{T}_2^*$.  Thus, a plausible signal rate that is greater than an order of magnitude faster than state-of-the-art transmon qubit decoherence times~\cite{girvin_circuit_2016,pfaff_controlled_2017,keller_transmon_2017} is $r = 2\pi \times 100$~kHz bandwith. For effective capture of this signal, we can choose a nominal value of $\kappa_e = 3r$. Since $\kappa_e \gg \kappa_i$, the coupling dynamics are dominated by the influence of $\kappa_e$.

We adopt the quantum input-output theory in the Markovian limit~\cite{gardiner1985input} and apply the Scattering-Hamiltonian-Lindbladian (SLH) formalism~\cite{combes_slh_2017}. This is justified by the fact that the input acoustic wave carries a single quantum of mechanical vibration, which is either lost to the large environment or transferred to the cavity as stored information. Since the loss rate from the cavity is much less than the oscillation frequency, it can be modeled as a nearly closed Hilbert space with only a perturbative coupling to the environment. In the Heisenberg picture, the master equation governing the time-evolution of the annihilation operator $a_c$ for the bosonic quantized field in the phononic cavity for this system is (see the SI for a detailed derivation):
\begin{equation} \label{Eq:master}
	\dot{a}_c = - \frac{1}{2} \left( 4 \kappa_e \cos^2 \left( \frac{\theta(t)}{2} \right) + \kappa_i \right) a_c - 2 \sqrt{\kappa_e} \cos \left( \frac{\theta(t)}{2} \right) a_\mathrm{in} - \sqrt{\kappa_i} f_i,
\end{equation}
where $\theta(t)$ represents the temporal profile of the phase shift induced by the principal phase shifter upon the rightward-traveling output, $-\theta(t)/2$ represents the profile of the phase shift induced by the IO phase shifter upon the overall (rightward-traveling) input and overall (leftward-traveling) output each, $a_\mathrm{in}$ is the input field, and $f_i$ is the vacuum input through the intrinsic loss channel. We also obtain the following input-output relation:
\begin{equation}
	a_\mathrm{out} (t) = a_\mathrm{in} (t) + 2 \sqrt{\kappa_e} \cos \left( \frac{\theta(t)}{2} \right) a_c.
\end{equation}
The system behaves as a single conventional resonant cavity with no scattering from input to output, and also possesses a real-valued raw output amplitude rate of $2 \kappa_e ( 1 + \cos (\theta(t)))$.

The fidelity is maximized by optimizing the phase profile $\theta(t)$ that is applied to the phase shifters. See the SI for a derivation of this optimal profile, which we calculate as the following:
\begin{align} \label{eq: theta(tau')}
\begin{split}
\theta(\tau') =
\begin{cases}
0, & 0 < \tau' < \tau'_c \\
\cos^{-1} \bigg(\frac{r}{2\kappa_e} e^{-\frac{r}{\kappa_e}\tau'} \Big(A_1 - A_2 e^{-\frac{r}{\kappa_e}\tau'}\Big)^{-1} - 1\bigg), & \tau' > \tau'_c
\end{cases},
\end{split}
\end{align}
where $\tau' = \kappa_e t$, and the constants $A_1$, $A_2$, and $\tau'_c$ are defined in the following manner:
\begin{align}
A_1 &= \frac{16}{\Big(4\sqrt{\frac{\kappa_e}{r}} + \sqrt{\frac{r}{\kappa_e}}\Big)^2} \bigg(\frac{8}{4 + \frac{r}{\kappa_e}}\bigg)^{-\frac{8}{4 - \frac{r}{\kappa_e}}} + \bigg(\frac{8}{4 + \frac{r}{\kappa_e}}\bigg)^{-\frac{2}{4\frac{\kappa_e}{r} - 1}}, \\
A_2 &= 1, \\
\tau'_c &= \frac{2}{4 - \frac{r}{\kappa_e}} \ln{\bigg(\frac{8}{4 + \frac{r}{\kappa_e}}\bigg)}.
\end{align}
The fidelity as a function of $r/\kappa_e$ is shown in \figureautorefname{ \ref{fig:phase_profile_and_fidelity}}(a). Given the ratio $r/\kappa_e = 1/3$, the optimal phase profile [\figureautorefname{ \ref{fig:phase_profile_and_fidelity}}(b)] yields a fidelity of 96.9\%. The rate of increase of the phase profile sets the scale for the necessary switching time of the phase shifter, while the lengths of the phase shifters dictate the voltage scale required on each one following \figureautorefname{ \Ref{fig:ps_op_pt}}(c). The maximum rate of increase for the phase is limited by the 14 MHz resonance in the transfer function [\figureautorefname{ \ref{fig:piezo_bias_effect}(d)], which corresponds to a rate of change $d\theta/d(\kappa_e t) \approx 23$ for $\kappa_e = 2\pi \times 300 \textrm{ kHz}$. We satisfy this condition by setting the time interval between consecutive discrete phase values such that the slope never exceeds the above value. In particular, the rise is sharpest immediately after the critical time $t_c$ is passed. Here, we find that setting the next point in time as $1.1\kappa_e t_c$ keeps the slope within the physically achievable range. This corresponds to a time interval of $0.1\kappa_e t_c$. A full-quantum simulation on QuTiP (Quantum Toolbox in Python)~\cite{johansson2012qutip} reveals that the fidelity remains at 96.9\% (equivalent to the value calculated for continuous time) even with this interval, implying that the time steps are sufficiently Riemann-like. However, this idealized fidelity calculation does not account for the finite time that it takes for a signal to propagate through the principal phase shifter and return to the cavity. During this interim, the population of the cavity changes, thereby disturbing the carefully designed destructive interference between the leftward-traveling input and the leftward-traveling raw output and resulting in increased loss (see the SI for a detailed derivation of the time-evolution of the cavity operator in the presence of delay). For a given round trip time $\delta_F$ from the cavity to the mirror and back, a time lag can be introduced to the phase profile of the mirror as well as to the temporal profile of the cavity detuning. We label these time delays as $\delta_M$ and $\delta_C$, respectively. For $\delta_F \approx 60 \textrm{ ns}$, \figureautorefname{ \ref{fig:phase_profile_and_fidelity}}(c) depicts the results of a numerical simulation calculating the maximum fidelity as a function of $\delta_M$, along with the corresponding value of $\delta_C$. The maximum fidelity approximately equals 89.0\%, attained when $\delta_M = 21 \textrm{ ns}$ and $\delta_C = -34 \textrm{ ns}$.
\begin{figure*}[!tb]
	\captionsetup[subfigure]{justification=centering}
	\centering
	\subfloat[Fidelity as a function of $r/\kappa_e$]{\includegraphics[width=0.5\linewidth]{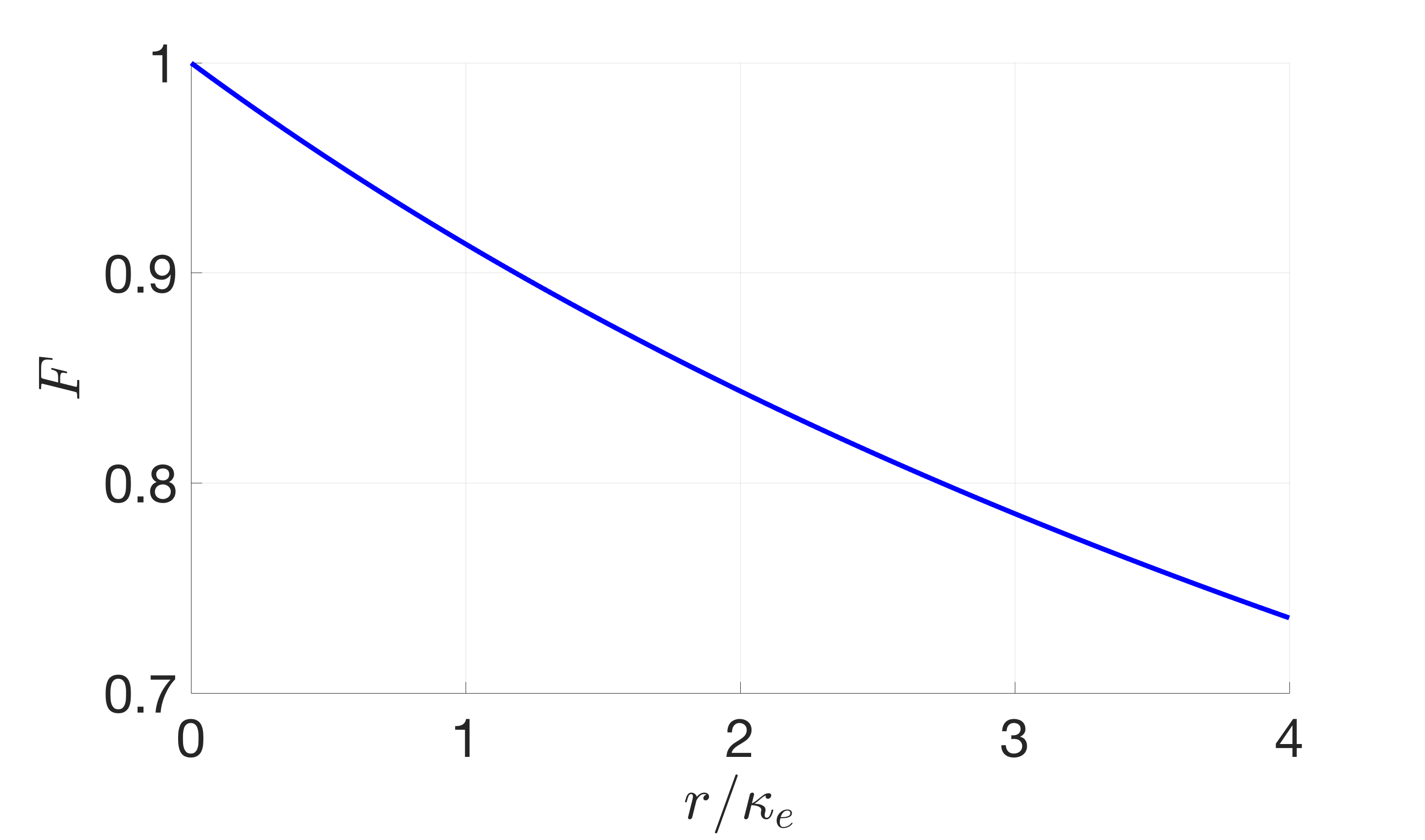}}
	\subfloat[$\theta(t)$, with time in units of $1/\kappa_e$, given $r = \kappa_e/3$]{\includegraphics[width=0.5\linewidth]{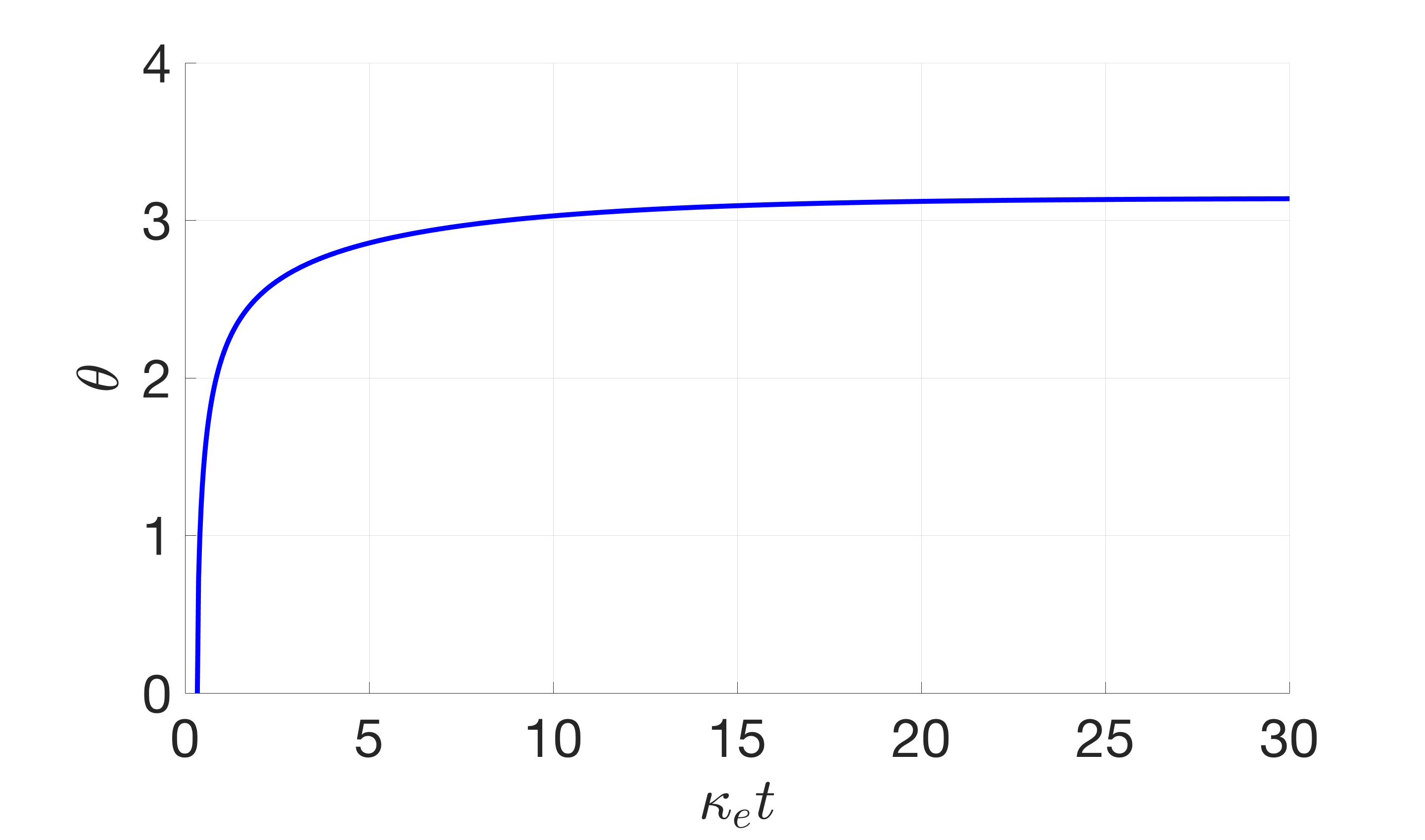}}
	\\
	\subfloat[Maximum fidelity as a function of $\delta_M$, and the corresponding $\delta_C$]{\includegraphics[width=0.5\linewidth]{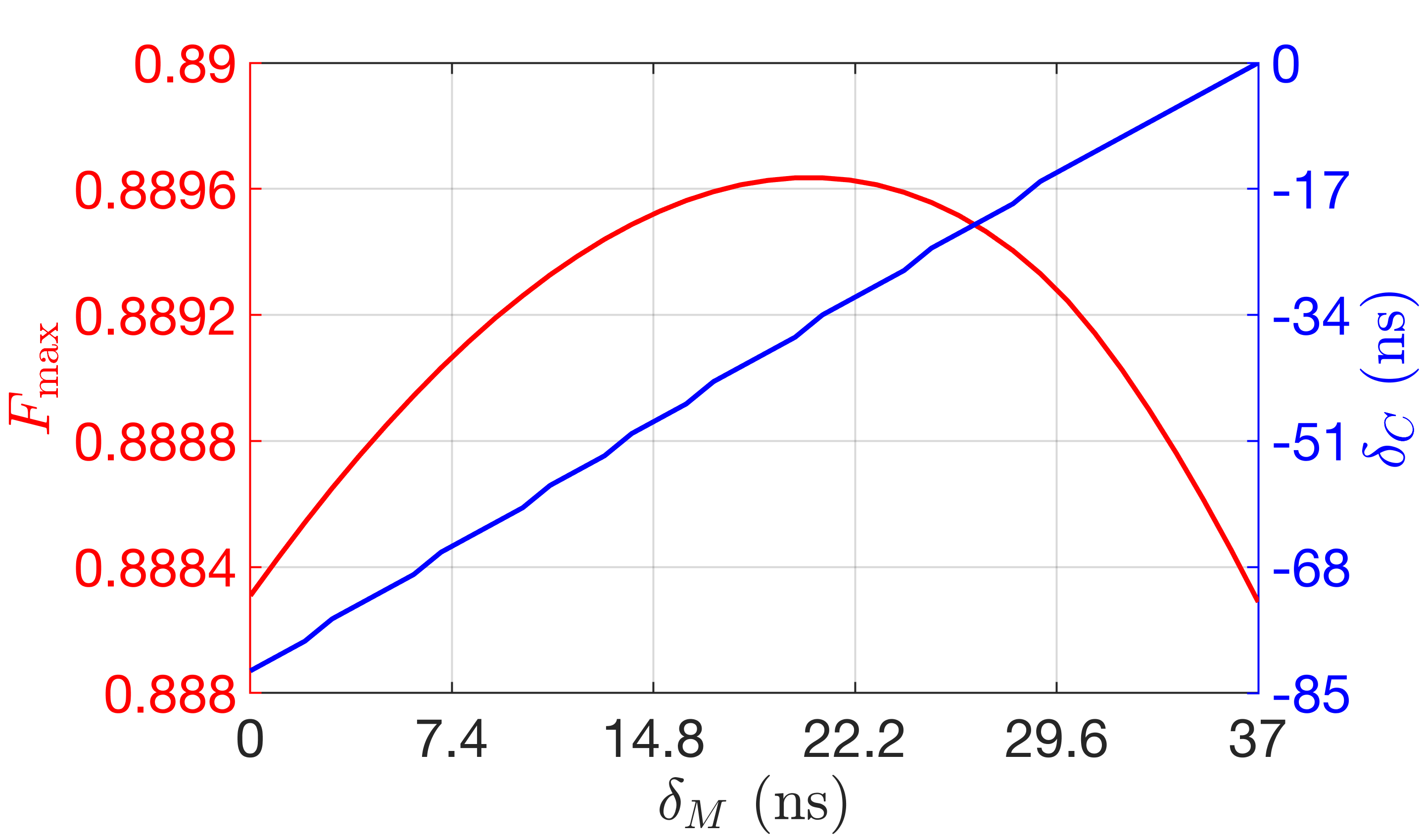}}
	\caption{Simulation result of the phononic cavity with tunable coupling. The simulation parameters can be found in the main text.}
    \label{fig:phase_profile_and_fidelity}
\end{figure*}

Because the signal must pass by the cavity and reflect back to the cavity before the interference effect can take hold, we must minimize the time cost associated with instantiating the interference in order to maximize the fidelity of the transfer. Therefore, in our suggested device realization shown in \figureautorefname{ \Ref{fig:Phononic_memory_architecture}}(c)  the more capable tensile operating sense is employed for the principal phase shifter, which extends for only 10 periods. The cladding surrounding the cavity extends for 7 periods, and there is an additional 1 period separating the phase shifter and the cladding, so the total path length in the waveguide between the cavity and the mirror is 18 periods, or $9.54$ \textmu m. Since $v_g = 312$ m/s, the time delay $\delta_F$ associated with a signal passing from the cavity, through the principal phase shifter, reflecting at the mirror, and returning to the cavity is 61 ns. The voltage required to achieve a $+\pi$ phase shift in the principal phase shifter is approximately $-85$ V. On the other hand, since the time it takes for the signal to pass through the IO phase shifter does not affect the transfer fidelity of the memory, we relax the length requirement on it to 40 periods. Consequently, the bias required for a $-\pi$ phase shift upon passing twice through the input phase shifter is $+35$ V.

Improving the fidelity of the device specified in \figureautorefname{ \Ref{fig:Phononic_memory_architecture}}(c) above 90\% is immediately possible by shortening the length of the principal phase shifter, necessitating an increased bias on its piezoelectric actuators. Since $\delta_F$ determines the fidelity limit, additional improvements could be made by identifying a waveguide geometry that yields a larger phase shift at a higher group velocity to decrease the time cost associated with trapping the qubit in the cavity. Decreasing the operating voltage would improve the compatibility with the cryogenic environment in which the system will exist. We hypothesize that further study of the scaling of the acoustoelastic effect with decreasing feature size could yield an augmented frequency shift. Smaller feature sizes would likely increase the operating frequency and also place more stringent the demands on the electron beam lithography process required to fabricate the PnC and waveguide.

\subsection{Programmable Multi-Mode Phonon Inteferometers and Addressable Memory Registers}

In photonic information processing, Mach-Zehnder interferometers (MZI) are  critical building blocks for both classical and quantum applications because they can be multiplexed together to provide an SU$(N)$ transformation from $N$ input spatial modes to $N$ output modes~\cite{harris_linear_2018}. The same is true for phononic processors. \figureautorefname{ \ref{fig:MZ_bias}}(a) shows an SU$(2)$ transformation
\begin{equation}
    U(2)=i\begin{pmatrix}
e^{i\phi/2}\sin{\theta/2} & e^{i\phi/2}\cos{\theta/2} \\
e^{-i\phi/2}\cos{\theta/2} & -e^{-i\phi/2}\sin{\theta/2} 
\end{pmatrix}
\end{equation}
that would be achieved by arranging the previous section's phase shifters in differentially biased pairs both between ($\theta$) and after ($\phi$) a pair of 50:50 beam splitters and combiners. The mode-combining and mode-splitting directional couplers convert input modes $a_1$ and $a_2$ into output modes $b_{1,2}=(1/\sqrt{2})(a_{1,2}+ia_{2,1})$ and can be realized by evanescently coupling the phononic crystal waveguides~\cite{olsson_iii_microfabricated_2008}.
\begin{figure}
	\centering
	\includegraphics[width=\textwidth]{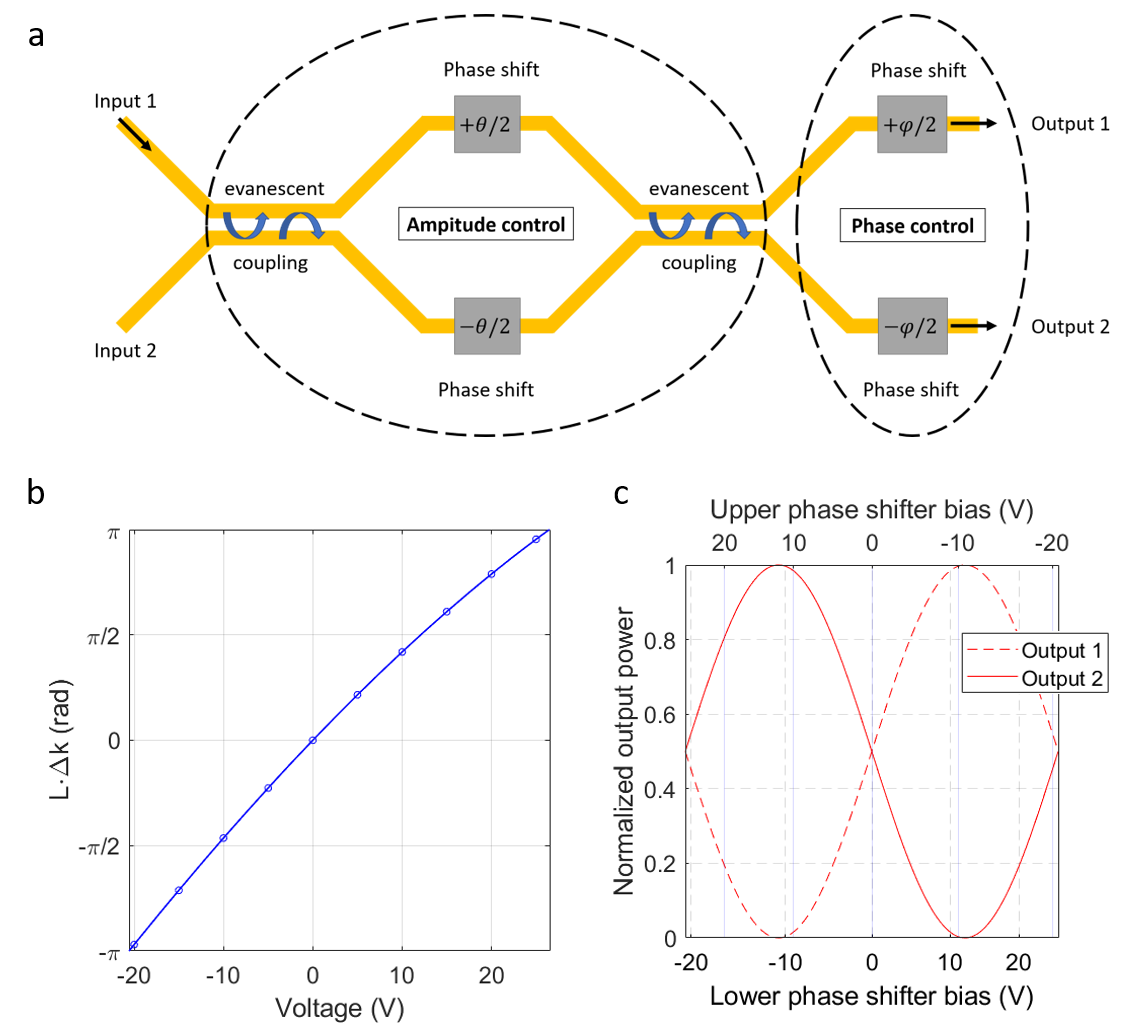}
	\caption{(a) Schematic diagram of an SU$(2)$ phononic Mach-Zehnder interferometer switch. (b) Phase shift $L\cdot\Delta k$ versus voltage $V$ for a 100 period phase shifter operated at 5.1406 GHz with $k_x=0.6897\pi/a$. (c) Output power as a function of phase shifter bias for the Mach-Zehnder interferometer.}
	\label{fig:MZ_bias}
\end{figure}

For specificity, we assume that the phase shifters in this structure are 100 periods long, and that the same 5.1406 GHz operating frequency is employed as was used in the above sections. \figureautorefname{ \ref{fig:MZ_bias}}(b), shows the relationship between voltage and phase for a single phase shifter, using the data from \figureautorefname{ \ref{fig:ps_op_pt}}. As an example of how the 2-mode interferometer operates, consider the case where an incoming signal is present in only one of the input channels of the interferometer. The power in the two interferometer outputs is a function of the phase difference $\theta$ between the two paths and is proportional to $(1\pm\sin{\theta})/2$. The bias levels on the phase shifters are specified on the abscissas in \figureautorefname{ \ref{fig:MZ_bias}}(c), showing that the complete range of interference between the channels is accessible at less than $\pm 25$ V. 

\begin{figure}
	\centering
	\includegraphics[width=\textwidth]{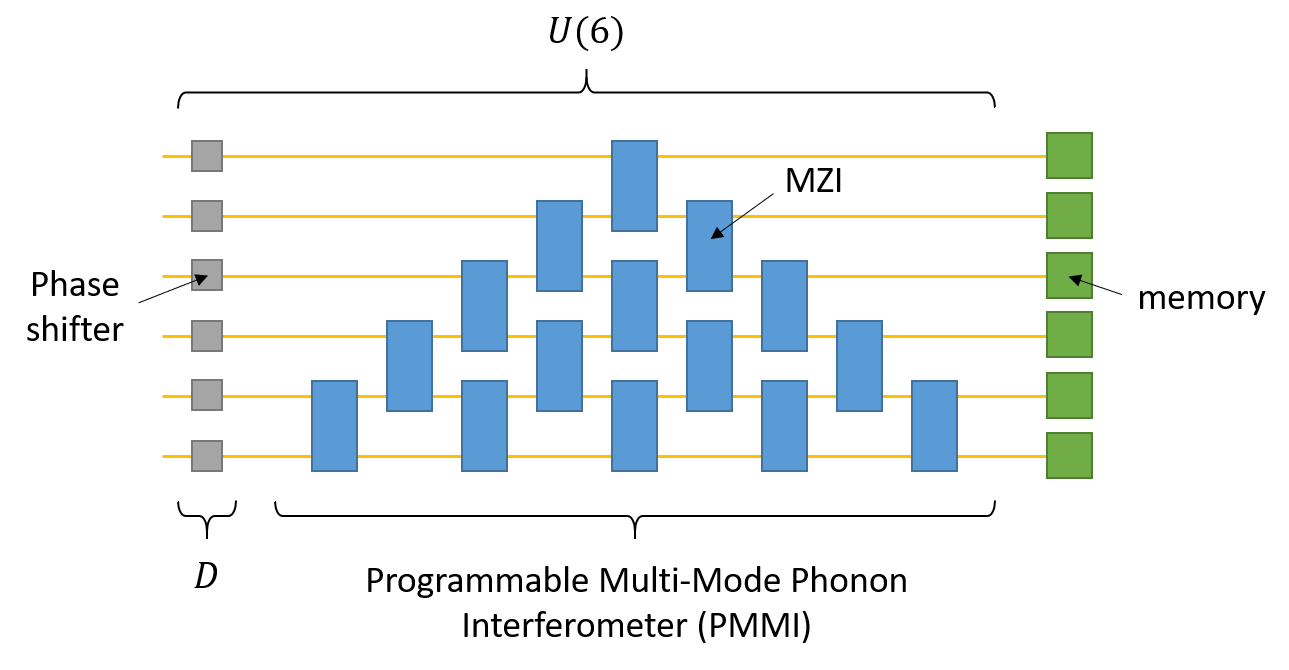}
	\caption{Schematic of an addressable phononic memory constructed from a PMMI terminated with an array of cavities that can store quantum phononic information. The phase screening elements in $D$, shown as gray boxes, represent single mode phase shifters [\figureautorefname{ \ref{fig:phase_shifter_structure}}], while the blue boxes represent the $U(2)$ MZI elements [\figureautorefname{ \ref{fig:MZ_bias}}] that make up the PMMI, and the green boxes represent cavities with tunable coupling [\figureautorefname{ \ref{fig:Phononic_memory_architecture}}].}
	\label{fig:MZ_tree}
\end{figure}

Cascading networks of these SU$(2)$ transformations enable programmable multi-mode interferometers (PMMIs) to be created that implement SU$(N)$ transformations between $N$ inputs and $N$ outputs~\cite{harris_linear_2018}. These piezoelectrically actuated, acoustomechanically tuned phononic PMMIs would be highly analogous to recently demonstrated piezoelectrically actuated, optomechanically tuned photonic PMMIs~\cite{dong_high-speed_2021}. Combining these PMMI-enabled SU$(N)$ transformations with non-classical phononic sources such as Fock states or squeezed states would enable universal linear phononic quantum processing, analogous to univeral linear quantum optical processors~\cite{carolan_universal_2015}. Finally, by combining these PMMI-enabled SU$(N)$ transformations with the quantum phononic memory elements described above, an addressable quantum memory register can be created. In this way, $M$ phononic modes can be coupled with arbitrary amplitudes and phases to $N$ quantum phononic memory qubits, where $M\leq N$. We sketch one possible network structure in \figureautorefname{ \ref{fig:MZ_tree}}, where 6 phononic qubit input modes can be routed into any one of 6 memories through a 6-mode PMMI using the decomposition algorithm from Ref.~\cite{reck_experimental_1994}. The phase screening element $D$ preceding the PMMI consisists of phase shifters on each of the input ports and is required to realize any arbitrary transformation in $U(N)$~\cite{harris_linear_2018}. An alternative formulation of the phase screen is possible such that it follows the PMMI transformation~\cite{reck_experimental_1994,clements_optimal_2016}. 


\section{Conclusions}

We have proposed harnessing piezo-acoustomechanical interactions in non-piezoelectric materials induced by strain actuators to engineer a promising class of phononic phase shifting devices. Quantum information processing elements such as programmable multi-mode interferometers, reconfigurable quantum memories, switches, and tunable mirrors can be constructed by integrating phase shifting elements into PnC waveguide circuits. Although we have focused on a particular set of materials, Sc$_x$Al$_{1-x}$N films deposited on SOI wafers, the design concepts we have presented are also applicable to LiNbO$_3$, diamond, GaAs, and other low loss acoustic materials that are amenable to phononic information processing~\cite{wu_microwave--optical_2020}. For example, incorporating the nonlinear piezoelectric effect~\cite{branch_investigation_2020} could aid in the design of quantum phononic components in LiNO$_3$ and GaAs, both of which benefit from prior studies regarding of higher order constitutive properties~\cite{cho_nonlinear_1987,mcskimin_thirdorder_1967}. The $\overbar{4}3m$ point group of GaAs also shares the same symmetry reduction in the third order elastic moduli as the $m\overbar{3}m$ point group of Si~\cite{hearmon_third-order_1953}, so the elastic constitutive model applied here is directly transferred to GaAs. While this paper has focused on using quasi-static strain with piezoelectric transducers to produce the requisite modulation of phonon phase velocity, it may also be possible to use microwave frequency control fields and intrinsic acoustic nonlinearities (analogous to $\chi^{(2)}$ and $\chi^{(3)}$ optical nonlinearities) to dynamically modulate cavity coupling rates and reconfigure circuits \cite{heuck_controlled-phase_2020,heuck_photon-photon_2020}. We anticipate that further exploration of acoustomechanical interactions in these material platforms will yield a rich variety of quantum information processing devices that have the potential to form the building blocks of a new generation of quantum hardware, where quantum information processing and storage takes place in the phononic domain.

\section{Acknowledgements}

This work was partially funded by the U.S. Department of Energy, Office of Science, Advanced Scientific Computing Research (ASCR) under FWP 19-022266 `Quantum Transduction and Buffering Between Microwave Quantum Information Systems and Flying Optical Photons in Fibers'. In addition, partial support was provided by the Laboratory Directed Research and Development program at Sandia National Laboratories, a multimission laboratory managed and operated by National Technology and Engineering Solutions of Sandia, LLC., a wholly owned subsidiary of Honeywell International, Inc., for the U.S. Department of Energy's National Nuclear Security Administration under Contract No. DE-NA-003525. M.E. performed this work, in part, at the Center for Integrated Nanotechnologies, an Office of Science User Facility operated for the U.S. Department of Energy (DOE) Office of Science. This paper describes objective technical results and analysis. Any subjective views or opinions that might be expressed in the paper do not necessarily represent the views of the U.S. Department of Energy or the United States Government.

\widetext
\clearpage
\begin{center}
\textbf{\large Supplemental Information}
\end{center}
\setcounter{equation}{0}
\setcounter{figure}{0}
\setcounter{table}{0}
\setcounter{section}{0}
\makeatletter

\renewcommand{\thefigure}{S\arabic{figure}}
\renewcommand{\thetable}{S\Roman{table}}
\renewcommand{\thesection}{S\Roman{section}}
\renewcommand{\theequation}{S\arabic{equation}}

\section{Hyperelastic constitutive model for the acoustoelastic effect in silicon}

Due to the relatively small defect density in its covalent network, diamond cubic Si does not readily undergo plastic deformation when in a state of high strain. Absent exotic external conditions, such as the high compression which causes Si to transform to a metallic state~\cite{gridneva_phase_1972} or a hydrostatic deformation that can counteract its brittle nature~\cite{rabier_plastic_2007}, Si predominantly undergoes elastic deformation prior to fracture. Moreover, nanoscale structures fabricated out of Si can achieve such high strains that the infinitesimal strain limit of linear elasticity theory is grossly violated~\cite{walavalkar_controllable_2010}. In rectangular cartesian coordinates, the components of the strain field $s_{ij}$ are defined in terms of the position derivatives of the displacement field $\mathbf{u}$:
\begin{equation} \label{Eq:strain}
    s_{ij} = \frac{1}{2}\left(\frac{\partial u_i}{\partial r_j} + \frac{\partial u_j}{\partial r_i} + \frac{\partial u_k}{\partial r_i}\frac{\partial u_k}{\partial r_j} \right)
\end{equation}
where $r_i$ are the components of the equilibrium position vector.
In the infinitesimal theory, the quadratic terms in Eq. \ref{Eq:strain} are dropped, yielding a linearized strain-displacement relation that can be redefined in terms of the symmetric gradient~\cite{auld_acoustic_1990}. We include the geometric nonlinearity of the system by retaining these higher order terms, and in addition employ a hyperelastic material model that extends the Taylor series expansion of the elastic strain energy density function $W$ for an anisotropic single crystal to third order~\cite{brugger_thermodynamic_1964}:
\begin{equation} \label{Eq:Uexpansion}
    W = W_0 + W_1 + W_2 + W_3 + \cdots
\end{equation}
In general, a hyperelastic material is one for which the scalar elastic potential energy function $W$ exists so that the stress tensor can be defined (at constant entropy) by $T_{ij}=\partial W / \partial s_{ij}$~\cite{ogden_non-linear_1997,royer_elastic_2000}. Neither $W_0$ nor $W_1$ contribute since the expansion is taken about the state of zero strain. The elasticity tensor is the fourth-order tensor $c_{ijkl}$ that relates stress and strain following $T_{ij} = c_{ijkl} s_{kl}$ when Eq. \ref{Eq:Uexpansion} is terminated at second order. By retaining $W_3$ a strain-dependent correction to the elasticity tensor (\textit{i.e.} the `phonoelastic tensor') is added. To yield the complete acoustoelastic change in propagation constant $k$ in a finite element simulation, the constitutive relation $W=W_2 + W_3$ can be employed, thus incorporating the acoustoelastic effect into the acoustic field equations, the solution of which naturally incorporates the moving boundary effect.

The hyperelastic material model we use to define the constitutive elasticity relation for Si is derived following the convention used to report the second and third order elastic moduli of Si that we employ in our calculations~\cite{hall_electronic_1967}. The second order term in \equationautorefname{ \ref{Eq:Uexpansion}} is defined as
\begin{equation} \label{Eq:2ndorder}
	W_{2} =\frac{1}{2} c_{IJ} s_{I} s_{J}=\frac{1}{2} \sum_J c_{JJ} s_J^2 + \sum_{J<P} c_{JP} s_J s_P,
\end{equation}
where Voigt notation has been used to reindex the 2nd order elastic moduli and the components of the strain field: $c_{ijkl} \rightarrow c_{IJ}$ and $s_{ij} \rightarrow s_I$~\cite{nye_physical_1984}. 
To incorporate the acoustoelastic effect, we include the third order term in the series expansion of the strain energy density,
\begin{equation} \label{Eq:3rdorder}
	W_3= \frac{1}{6} c_{IJK} s_I s_J s_K = \frac{1}{6} \sum_J c_{JJJ} s_J^3 +\frac{1}{2} \sum_{J \neq P} c_{JJP}  s_J^2 s_P +\sum_{J<P<R} c_{JPR} s_J s_P s_R,
\end{equation}
which augments the elasticity tensor with a strain-dependent component. Thus, the acoustoelastic effect results from a sixth order tensor (the 6 indices of which are reindexed down to 3 in \equationautorefname{ \ref{Eq:3rdorder}}) that contracts the third order strains in the expansion of the scalar elastic strain energy density~\cite{brugger_thermodynamic_1964}. The acoustoelastic effect modifies the speed of sound as a function of strain and is similar to the more familiar photoelastic effect, in which strain causes a change in the dielectric constant of a solid. However, the tensorial description of the effects differ since the photoelastic tensor is a fourth order tensor that modifies the principal components of the second order relative dielectric permeability tensor, which at optical frequencies represent the inverse principal refractive indices~\cite{nye_physical_1984}.

\subsection{Elastic moduli}

The point group for Si is $m\bar{3}m$, which yields a significant reduction in the number of independent elastic moduli~\cite{hearmon_third-order_1953}. At second order, the three independent moduli are $c_{11}, c_{12}$, and $c_{44}$ because the following relations hold and all other terms are zero:
\begin{eqnarray}
\begin{aligned}
&c_{11}=c_{22}=c_{33}, \nonumber \\
&c_{12}=c_{21}=c_{13}=c_{31}=c_{23}=c_{32}, \nonumber \\
&c_{44}=c_{55}=c_{66}. \nonumber \\
\end{aligned}
\end{eqnarray}
The crystal symmetry also yields a dramatic reduction in the number of independent third order elastic moduli. The six independent tensor elements $c_{111}, c_{112}, c_{144}, c_{166}, c_{123}$, and $c_{456}$ can be related to the other nonzero components as follows:
\begin{eqnarray}
\begin{aligned}
    &c_{111}=c_{222}, \nonumber \\ &c_{112}=c_{113}=c_{122}=c_{133}=c_{223}=c_{233}=c_{333}, \nonumber \\
    &c_{144}=c_{255}=c_{366}, \nonumber \\ &c_{166}=c_{155}=c_{244}=c_{266}=c_{355}=c_{344}. \nonumber \\
\end{aligned}
\end{eqnarray}
For the simulations in this work, the elastic moduli reported in Ref.~\cite{hall_electronic_1967} are employed, and are listed in \tableautorefname{ \ref{tab:moduli}}.
\begin{table}[h]
    \centering
	\begin{tabular}{c|c}
		Modulus & Value (GPa) \\
		\hline
		$c_{11}$ & 165.64 \\
		$c_{12}$ & 63.94 \\
		$c_{44}$ & 79.51 \\
		$c_{111}$ & -795 \\
		$c_{112}$ & -445 \\
		$c_{123}$ & -75 \\
		$c_{144}$ & 15 \\
		$c_{166}$ & -310 \\
		$c_{456}$ & -86 \\
		\hline
	\end{tabular}
	\caption{Second and third order elastic moduli for Si from Ref.~\cite{hall_electronic_1967}.}
	\label{tab:moduli}
\end{table}
Stated in terms of the reduced set of elastic moduli, the second and third order elastic strain energies can be written following \equationautorefname{ \ref{Eq:2ndorder} and \ref{Eq:3rdorder}}:
\begin{eqnarray}
	W_2 &=& \frac{1}{2} c_{11} (s_1^2+s_2^2+s_3^2 )+c_{12} (s_1 s_2+s_1 s_3+s_2 s_3)+  \frac{1}{2} c_{44} (s_4^2+s_5^2+s_6^2 ), \\
	W_3 &=& \frac{1}{6} c_{111} (s_1^3+s_2^3+s_3^3 )+\frac{1}{2} c_{112} (s_1^2 s_2+s_1^2 s_3+s_1 s_2^2+s_1 s_3^2+s_2^2 s_3+s_2 s_3^2 ) \nonumber \\
	&& +\frac{1}{2} c_{144} (s_1 s_4^2+s_2 s_5^2+s_3 s_6^2 )+\frac{1}{2} c_{166} (s_1 s_5^2+s_1 s_6^2+s_2 s_4^2+s_2 s_6^2+s_3 s_4^2+s_3 s_5^2 )\nonumber \\
	&&+c_{123} s_1 s_2 s_3+c_{456} s_4 s_5 s_6.
\end{eqnarray}

\subsection{COMSOL implementation}

In COMSOL, it is possible to implement a user-defined hyperelastic material defined by an elastic strain energy function $W$. The acoustoelastic effect can therefore be included by setting $W=W_2+W_3$. Validation of this approach was provided by the equivalence between results obtained from computations in which $W=W_2$ and computations in which a linear elastic material was employed with a geometric nonlinearity. Our devices were oriented such that the waveguide propagation direction was parallel to the [110] direction in the wafer. This orientation was specified by transforming the material frame strain variables referenced to COMSOL's cartesian axes (here denoted with a prime $s'_{ij}$) so that they could be used to define the Voigt strain variables $s_I$ in terms of the [100] orientation in which the elastic strain energy is conveniently defined. The transformation of variables $[s_{[100]} ]=[a]^T [s_{[110]} ][a]$ was performed using the transformation matrix
\begin{equation}
	[a] = \left[ \begin{array}{ccc} 
		1/\sqrt{2} & 1/\sqrt{2} & 0 \\
	-1/\sqrt{2} & 1/\sqrt{2} & 0 \\
	0 & 0 & 1 \end{array} \right],
\end{equation}
and the resulting expressions for the [100] Si Voigt strains stated in terms of the [110] strains are
\begin{eqnarray}
	s_1 &=& \frac{1}{2} s'_{xx} - s'_{xy} + \frac{1}{2} s'_{yy}, \nonumber \\
	s_2 &=& \frac{1}{2} s'_{xx} + s'_{xy} + \frac{1}{2} s'_{yy}, \nonumber \\
	s_3 &=& s'_{zz}, \nonumber \\
	s_4 &=& \sqrt{2} (s'_{xz} + s'_{yz}), \nonumber \\
	s_5 &=& \sqrt{2} (s'_{xz} - s'_{yz}), \nonumber \\
	s_6 &=& s'_{xx} - s'_{yy}. \nonumber
\end{eqnarray}

\subsection{Phonoelastic tensor}

The components of the phonoelastic tensor can be derived by solving $T_I=\partial W/\partial s_I = \tilde{c}_{IJ}s_J$ subject to the constraint $\tilde{c}_{IJ} = \tilde{c}_{JI}$. The results of this factorization are listed below. The tensor elements capped with a tilde $\tilde{c}_{IJ}$ include both the second order terms (consistent with the conventional definition of the elasticity tensor) and the third order phonoelastic components for the [100] orientation.
\begin{eqnarray}
	\tilde{c}_{11} &=& c_{11}+\frac{1}{2} c_{111} s_1+ \frac{1}{2} c_{112} (s_2+s_3), \nonumber \\
	\tilde{c}_{12} &=& c_{12} + \frac{1}{2} c_{112} (s_1 + s_2) + \frac{1}{2} c_{123} s_3, \nonumber \\
	\tilde{c}_{13} &=& c_{12} + \frac{1}{2} c_{112} (s_1 + s_3) + \frac{1}{2} c_{123} s_2, \nonumber \\
	\tilde{c}_{14} &=& \frac{1}{2} c_{144} s_4, \nonumber \\
	\tilde{c}_{15} &=& \frac{1}{2} c_{166} s_5, \nonumber \\
	\tilde{c}_{16} &=& \frac{1}{2} c_{166} s_6, \nonumber \\
	\tilde{c}_{22} &=& c_{11} + \frac{1}{2} c_{111} s_2 + \frac{1}{2} c_{112} (s_1 + s_3), \nonumber \\
	\tilde{c}_{23} &=& c_{12} + \frac{1}{2} c_{112} (s_2 + s_3) + \frac{1}{2} c_{123} s_1, \nonumber \\
	\tilde{c}_{24} &=& \frac{1}{2} c_{166} s_4, \nonumber \\
	\tilde{c}_{25} &=& \frac{1}{2} c_{144} s_5, \nonumber \\
	\tilde{c}_{26} &=& \frac{1}{2} c_{166} s_6, \nonumber \\
	\tilde{c}_{33} &=& c_{11} + \frac{1}{2} c_{111} s_3 + \frac{1}{2} c_{112} (s_1 + s_2), \nonumber \\
	\tilde{c}_{34} &=& \frac{1}{2} c_{166} s_4, \nonumber \\
	\tilde{c}_{35} &=& \frac{1}{2} c_{166} s_5, \nonumber \\
	\tilde{c}_{36} &=& \frac{1}{2} c_{144} s_6, \nonumber \\
	\tilde{c}_{44} &=& c_{44} + \frac{1}{2} c_{144} s_1 + \frac{1}{2} c_{166} (s_2 + s_3), \nonumber \\
	\tilde{c}_{45} &=& \frac{1}{2} c_{456} s_6, \nonumber \\
	\tilde{c}_{46} &=& \frac{1}{2} c_{456} s_5, \nonumber \\
	\tilde{c}_{55} &=& c_{44} + \frac{1}{2} c_{144} s_2 + \frac{1}{2} c_{166} (s_1 + s_3), \nonumber \\	
	\tilde{c}_{56} &=& \frac{1}{2} c_{456} s_4, \nonumber \\
	\tilde{c}_{66} &=& c_{44} + \frac{1}{2} c_{144} s_3 + \frac{1}{2} c_{166} (s_1 + s_2). \nonumber
\end{eqnarray}
The Bond method can be used to transform these components $\tilde{c}_{IJ}$ for the [100] orientation into components $\tilde{c}'_{IJ}$ for the [110] orientation by applying the matrix transformation $[ c_{[110]} ]=[M][c_{[100]} ][M]^T$. Here $[M]$ describes a rotation by an arbitrary angle $\xi$ about the [001] axis. Thus, rotating the [100] compnents into the [110] components requires setting the rotation angle $\xi = \pi/4$:
\begin{equation}
	[M] = \left[ \begin{array}{cccccc} 
		\cos^2\xi & \sin^2 \xi & 0 & 0 & 0 & \sin 2 \xi \\
		\sin^2 \xi & \cos^2 \xi & 0 & 0 & 0 & - \sin 2 \xi \\
		0 & 0 & 1 & 0 & 0 & 0 \\
		0 & 0 & 0 & \cos \xi & - \sin  \xi & 0 \\
		0 & 0 & 0 & \sin \xi & \cos \xi & 0 \\
		- \frac{\sin 2\xi}{2} & \frac{\sin 2 \xi}{2} & 0 & 0 & 0 & \cos 2 \xi
		\end{array} \right] = \left[ \begin{array}{cccccc} 
			\frac{1}{2} & \frac{1}{2} & 0 & 0 & 0 & 1 \\
			\frac{1}{2} & \frac{1}{2} & 0 & 0 & 0 & -1 \\
			0 & 0 & 1 & 0 & 0 & 0 \\
			0 & 0 & 0 & \frac{1}{\sqrt{2}} & - \frac{1}{\sqrt{2}} & 0 \\
			0 & 0 & 0 & \frac{1}{\sqrt{2}} & \frac{1}{\sqrt{2}} & 0 \\
			- \frac{1}{2} & \frac{1}{2} & 0 & 0 & 0 & 0
		\end{array} \right] .
\end{equation}
Finally, the modified elasticity tensor components for the [110] oriented system can be stated in terms of those for the [100] orientation as follows:
\begin{eqnarray}
	\tilde{c}'_{11} &=& \frac{1}{4} (\tilde{c}_{11} + \tilde{c}_{22}) + \frac{1}{2} \tilde{c}_{12} + \tilde{c}_{16} + \tilde{c}_{26} + \tilde{c}_{66}, \nonumber \\
	\tilde{c}'_{12} &=& \frac{1}{4} (\tilde{c}_{11} + \tilde{c}_{22}) + \frac{1}{2} \tilde{c}_{12} - \tilde{c}_{66}, \nonumber \\
	\tilde{c}'_{13} &=& \frac{1}{2} (\tilde{c}_{13} + \tilde{c}_{23}) + \tilde{c}_{36}, \nonumber \\
	\tilde{c}'_{14} &=& \frac{1}{\sqrt{2}} \left( \frac{1}{2} (\tilde{c}_{14} + \tilde{c}_{24} - \tilde{c}_{15} - \tilde{c}_{25})  +\tilde{c}_{46} - \tilde{c}_{56} \right), \nonumber \\
	\tilde{c}'_{15} &=& \frac{1}{\sqrt{2}} \left( \frac{1}{2} (\tilde{c}_{14} + \tilde{c}_{24} + \tilde{c}_{15} + \tilde{c}_{25})  +\tilde{c}_{46} + \tilde{c}_{56} \right), \nonumber \\
	\tilde{c}'_{16} &=& \frac{1}{2} (\tilde{c}_{26} - \tilde{c}_{16}) + \frac{1}{4} (\tilde{c}_{22} - \tilde{c}_{11}), \nonumber \\
	\tilde{c}'_{22} &=& \frac{1}{4} (\tilde{c}_{11} + \tilde{c}_{22}) + \frac{1}{2} \tilde{c}_{12} - \tilde{c}_{16} - \tilde{c}_{26} + \tilde{c}_{66}, \nonumber \\
	\tilde{c}'_{23} &=& \frac{1}{2} (\tilde{c}_{13} + \tilde{c}_{23}) - \tilde{c}_{36}, \nonumber \\
	\tilde{c}'_{24} &=& \frac{1}{\sqrt{2}} \left( \frac{1}{2} (\tilde{c}_{14} + \tilde{c}_{24} - \tilde{c}_{15} - \tilde{c}_{25})  -\tilde{c}_{46} + \tilde{c}_{56} \right), \nonumber \\
	\tilde{c}'_{25} &=& \frac{1}{\sqrt{2}} \left( \frac{1}{2} (\tilde{c}_{14} + \tilde{c}_{24} + \tilde{c}_{15} + \tilde{c}_{25})  - \tilde{c}_{46} - \tilde{c}_{56} \right), \nonumber \\
	\tilde{c}'_{26} &=& \frac{1}{2} (\tilde{c}_{16} - \tilde{c}_{26}) + \frac{1}{4} (\tilde{c}_{22} - \tilde{c}_{11}), \nonumber \\
	\tilde{c}'_{33} &=& \tilde{c}_{33}, \nonumber \\
	\tilde{c}'_{34} &=& \frac{1}{\sqrt{2}} (\tilde{c}_{34} - \tilde{c}_{35}), \nonumber \\
	\tilde{c}'_{35} &=& \frac{1}{\sqrt{2}} (\tilde{c}_{34} + \tilde{c}_{35}), \nonumber \\
	\tilde{c}'_{36} &=& \frac{1}{2} (\tilde{c}_{23} - \tilde{c}_{13}), \nonumber \\
	\tilde{c}'_{44} &=& \frac{1}{2} (\tilde{c}_{44} + \tilde{c}_{55}) - \tilde{c}_{45}, \nonumber \\
	\tilde{c}'_{45} &=& \frac{1}{2} (\tilde{c}_{44} - \tilde{c}_{55}), \nonumber \\
	\tilde{c}'_{46} &=& \frac{\sqrt{2}}{4} (-\tilde{c}_{14} + \tilde{c}_{24} + \tilde{c}_{15} - \tilde{c}_{25}), \nonumber \\
	\tilde{c}'_{55} &=& \frac{1}{2} (\tilde{c}_{44} + \tilde{c}_{55}) + \tilde{c}_{45}, \nonumber \\
	\tilde{c}'_{56} &=& \frac{\sqrt{2}}{4} (- \tilde{c}_{14} + \tilde{c}_{24} - \tilde{c}_{15} + \tilde{c}_{25}), \nonumber \\
	\tilde{c}'_{66} &=& \frac{1}{4} (\tilde{c}_{11} + \tilde{c}_{22}) - \frac{1}{2} \tilde{c}_{12}. \nonumber
\end{eqnarray}

\section{Computational Methodology}

\figureautorefname{ \ref{fig:comp_geom}} illustrates the geometry and methodology used to find the bandstructure and the phase shift induced from biasing the piezoelectric actuators. Bordering the suspended membrane region, 25 \textmu m of `bulk' material was added to approximate how the phase shifter would be situated within the die. A cross-sectional view of the simulation geometry is provided in \figureautorefname{ \ref{fig:comp_geom}}(a), illustrating this additional material as well as the interfaces at the extreme $y$- and $z$-coordinates far from the PnC where fixed constraints were placed. \figureautorefname{ \ref{fig:comp_geom}}(b) shows a close-up view of the PnC, how Floquet boundary conditions were placed along the propagation direction of the waveguide, as well as where $y$-symmetry conditions were enforced on the $xz$ plane bisecting the waveguide. An additional $z$-symmetry condition was placed on the $xy$ plane bisecting the waveguide for bandstructure computations, which were thus limited to the region highlighted in blue. The 4 combinations of $y$ and $z$ symmetric and anti-symmetric boundary conditions were calculated separately so that a fully symmetric waveguide mode could be identified. For bandstructure computations, the interface between the PnC and the overhanging Si membrane was fixed. The frequencies for the guided mode were selected and used as inputs for the ensuing computations that modeled the effect of piezo bias and therefore included the entire geometry. At zero piezo bias, the difference between the waveguide frequencies found in the limited bandstructure geometry and the full geometry were on the order of kHz. To compute the frequency shift at given values of bias $V$ and spatial periodicity $k_x$, the zero bias values from the full structure simulation were used.

For calculations involving the full structure, a central frequency of interest was identified from the zero bias bandstructure result and a basis size was specified such that the waveguide mode was captured when the full structure was relaxed at all biases of interest. Due to the frequency shift, a large enough window of frequency (i.e. bandwidth of the modal basis) is needed to capture the waveguide mode after it is shifted by $\Delta f$. However, the computation becomes increasingly expensive with a larger basis, so the basis size was increased iteratively for the bias that yielded the largest frequency shift, and then the modal displacement profiles of the solutions were viewed by eye to verify that the waveguide mode was captured in the subspace. Once a suitable basis size was identified, the full parametric sweep was initiated across the desired bias range. Subsequently, the modal solutions were analyzed by integrating the strain energy density $W$ in the waveguide region. The partitioned subset of the system used for the integration is highlighted in \figureautorefname{ \ref{fig:comp_geom}}(c), and the result of the integration for such a calculation is shown in \figureautorefname{ \ref{fig:comp_geom}}(d). For the guided mode, the strain energy in the waveguide was strongly peaked, by approximately 40 orders of magnitude, so identification of the guided mode frequencies according to the maximum value in each column of data was straightforward and unambiguous. All plots showing $\Delta f$ versus $V$ are derived by extracting the maximum values in each voltage column of a surface such as shown in \figureautorefname{ \ref{fig:comp_geom}}(d). The data plotted in Figure 4(b) and \figureautorefname{ \ref{fig:1.5GHz_results}}(c) are equivalent to the `top-down' view along the energy axis of a surface such as the one shown in \figureautorefname{ \ref{fig:comp_geom}}(d).

\begin{figure}
    \centering
    \includegraphics[width=\textwidth]{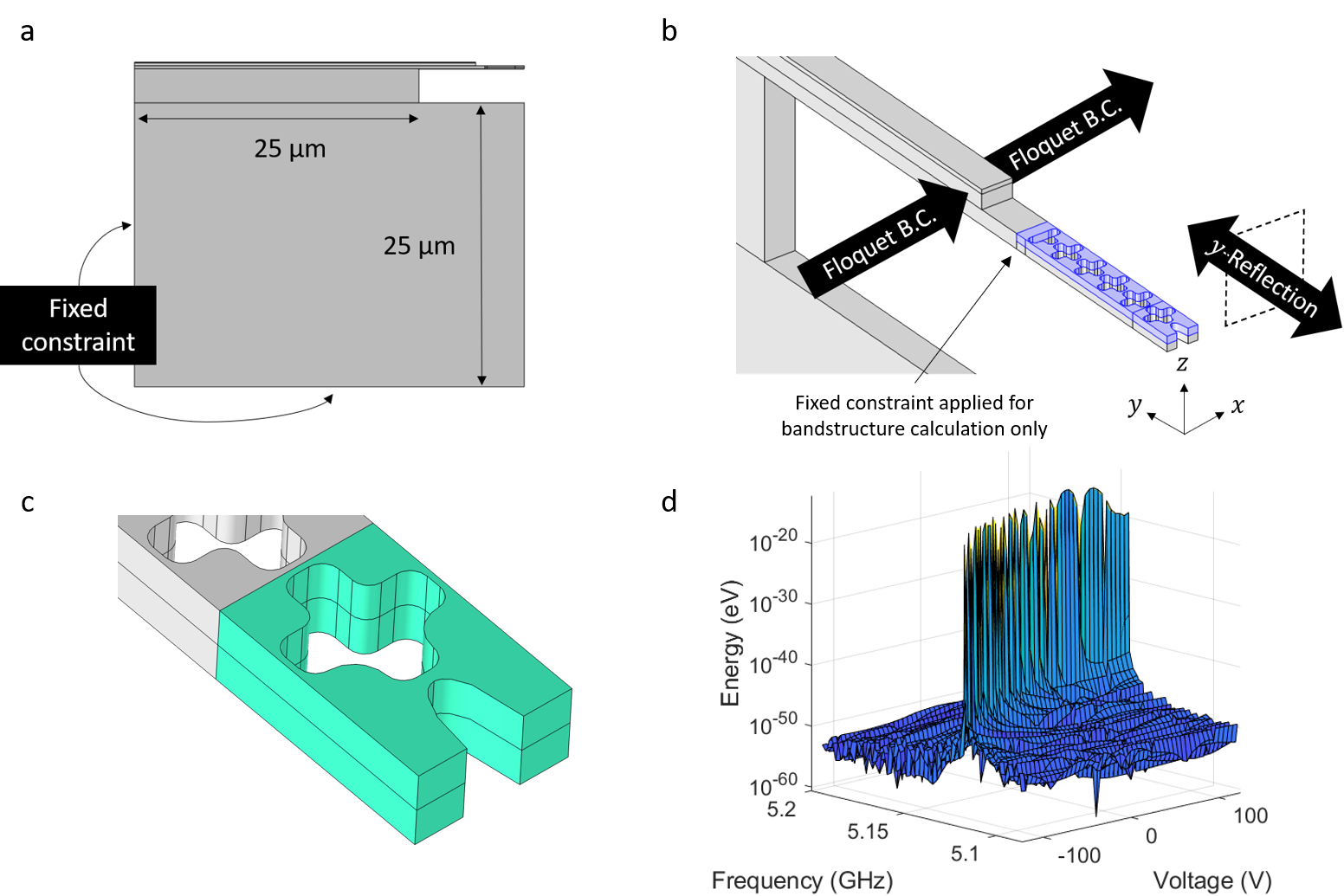}
    \caption{(a) Cross-sectional view of the structure used for FEM simulation. The system was extended by 25 \textmu m in height and depth. A fixed constraint was applied at the extreme $xy$ and $xz$ planes farthest from the membrane. (b) View of the suspended PnC region and piezoelectric actuator illustrating how Floquet boundary conditions and the symmetry plane through the middle of the waveguide were implemented. For bandstructure calculations, only the area highlighted in blue was included, and a fixed constraint was placed at the interfacial $xz$ plane where the bandstructure system contacts the remaining suspended membrane structure. (c) Close-up view of the waveguide. The region highlighted in turquoise was used for data analysis by integrating the strain energy function over this region to identify the waveguide mode in the basis of states. (d) Surface plot showing the sharply peaked strain energy in the guided mode, which is nearly 40 orders of magnitude above the strain energy that the vibrational modes in the bulk impart to the waveguide.}
    \label{fig:comp_geom}
\end{figure}

Frequency domain calculations of the transfer function were accomplished with continuous boundary conditions instead of Floquet boundary conditions imposed on the $yz$ planes bounding the structure. Transfer function plots represent the frequency dependence of the displacement of a point in the center of the waveguide. On the $xy$ plane that bisects the PnC, there are 5 spatial points defining the geometry of half the central ellipse in the waveguide. All of these points demonstrated the same frequency responses, and the frequency response of all nodal geometry points in the structure were plotted to verify that the points situated on the inner border of the ellipse in the center of the waveguide were representative.

We adopted the electro-acoustic tensor components for Sc$_{0.32}$Al$_{0.68}$N from Ref.~\cite{kurz_experimental_2019} except for $c^E_{12}$ which was calculated from Eq. 28 of Ref.~\cite{caro_piezoelectric_2015}. The $c$-axis of the piezoelectric material points up out of the surface of the film. We ignore the polycrystalline structure of the piezo due to the transverse isotropy of its hexagonal crystal structure. Since Al has low elastic anisotropy~\cite{cantwell_estimating_2012}, we treated it isotropically.

\section{PnC waveguide membrane with purely in-plane loading}

To demonstrate that the bowing in the deformation profile of the phase shifter arises from the placement of the piezoelectric actuators, and that the nonlinearity in the response of $\Delta f$ with $V$ arises from the bowing dynamics, a system with the piezo loading oriented purely along the $y$-axis in the plane transverse to the waveguide was constructed [\figureautorefname{ \ref{fig:y_load}}]. The same PnC and waveguide subsystem used for bandstructure computations in Figure 2 of the main text was employed. To create a piezo load purely along the $y$-axis, 1 \textmu m of Al$_{0.68}$Sc$_{0.32}$N placed at the back $xz$ interfacial plane of the PnC. The electro-acoustic tensor components were rotated such that the $c$-axis of the piezoelectric material was oriented along the $y$-axis of the geometry. Bias was placed on the $xz$ plane at the interface between the piezo and the silicon. The fixed back $xz$ plane of the piezo was grounded. As shown in \figureautorefname{ \ref{fig:y_load}}(b,c), the frequency shift of the system induced by piezo bias as well as the figure of merit $(V\cdot L)_\pi$ vary linearly with the piezo bias. Thus, if such a system could be fabricated where that the mechanical loading was entirely along the $y$-axis of the membrane, an ideal linear response could be obtained. The $z$-displacement profile of the biased system is plotted in \figureautorefname{ \ref{fig:y_load}}(d), showing that no bowing occurs. By contrast, the $z$-displacement for the system described in the main text [\figureautorefname{ \ref{fig:y_load}}(e)] shows that substantial bowing is induced. Thus, the bowing arises because of the loading geometry that results when the piezoelectric actuators are located on the top of the Si membrane.  

\begin{figure}
    \centering
    \includegraphics[width=\textwidth]{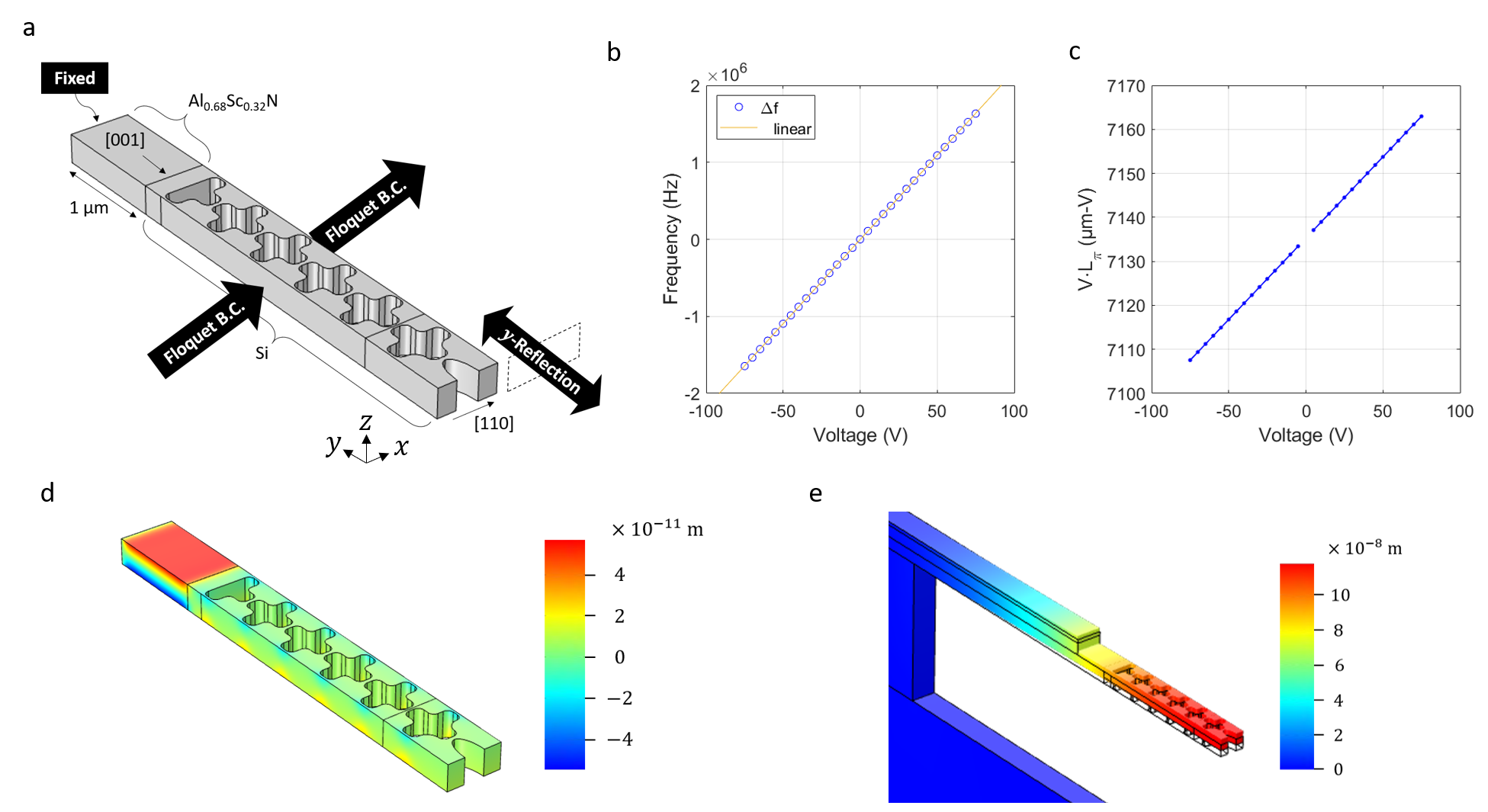}
    \caption{(a) Computational geometry with piezoelectric actuator oriented for in-plane loading of the membrane. The active $c$-axis of the Al$_{0.68}$Sc$_{0.32}$N crystal is oriented parallel to the $y$-axis of the membrane. (b) The linear relationship between frequency shift and piezo bias for the computational geometry shown in (a). (c) Demonstration that the $(V\cdot L)_\pi$ figure of merit accurately captures the response of a phase shifter with purely in-plane loading. (d,e) Comparison between the $z$-component of displacement  under a bias of $-50$ V for the in-plane loading structure and the structure described in the main text.}
    \label{fig:y_load}
\end{figure}

\section{PnC waveguide design with an operating frequency near 1.5 GHz}

An alternative waveguide geometry with a guided mode at 1.5 GHz was studied and serves as a foil for the 5 GHz structure described in the main text. The structure of the 1.5 GHz system was similar to the higher frequency system, except that it was scaled up by about 400\%. \figureautorefname{ \ref{fig:1.5GHz_structure}} displays the structure. The pitch of the PnC lattice was increased to $a=2.07$ \textmu m, with a unit cell geometry defined by paramaters $h=0.94a$, $w=0.2a$, and $r=w/2$. The structure was extended by 100 \textmu m from the membrane region, instead of the 25 \textmu m size of bulk material added to the 5 GHz system. An increased piezoelectric actuator overhang of 25 \textmu m was used, and the bare Si membrane between the piezo and the PnC was also increased to 5 \textmu m. The subset of the structure used for bandstructure calculations is highlighted in blue in \figureautorefname{ \ref{fig:1.5GHz_structure}}(a). The unit cell and the bandstructure of the infinitely periodic structure are displayed in  \figureautorefname{ \ref{fig:1.5GHz_structure}}(b). The size of the bandgap is reduced compared with the higher frequency structure. Note that the thickness of the device layer in the wafer remains the same, at 250 nm, for both the low and high frequency systems. The plan view geometry of the PnC and waveguide are displayed in \figureautorefname{ \ref{fig:1.5GHz_structure}}(c). As with the higher frequency system, the defect region spans 3 PnC lattice periods, and the outer radii of curvature for the crosses immediately adjacent to the defect are also increased to their maximum value of $r_d = (h - w)/2 - r$. The elliptical defect hole in the center of the waveguide has been converted to a circle with radius $h/2$. The PnC bandstructure for the waveguide system is shown in \figureautorefname{ \ref{fig:1.5GHz_structure}}(d), and the solutions are plotted according to their symmetry properties. The guided band that crosses 1.5 GHz is fully symmetric with respect to the $y$- and $z$-axes bisecting the waveguide.

\begin{figure}
    \centering
    \includegraphics[width=\textwidth]{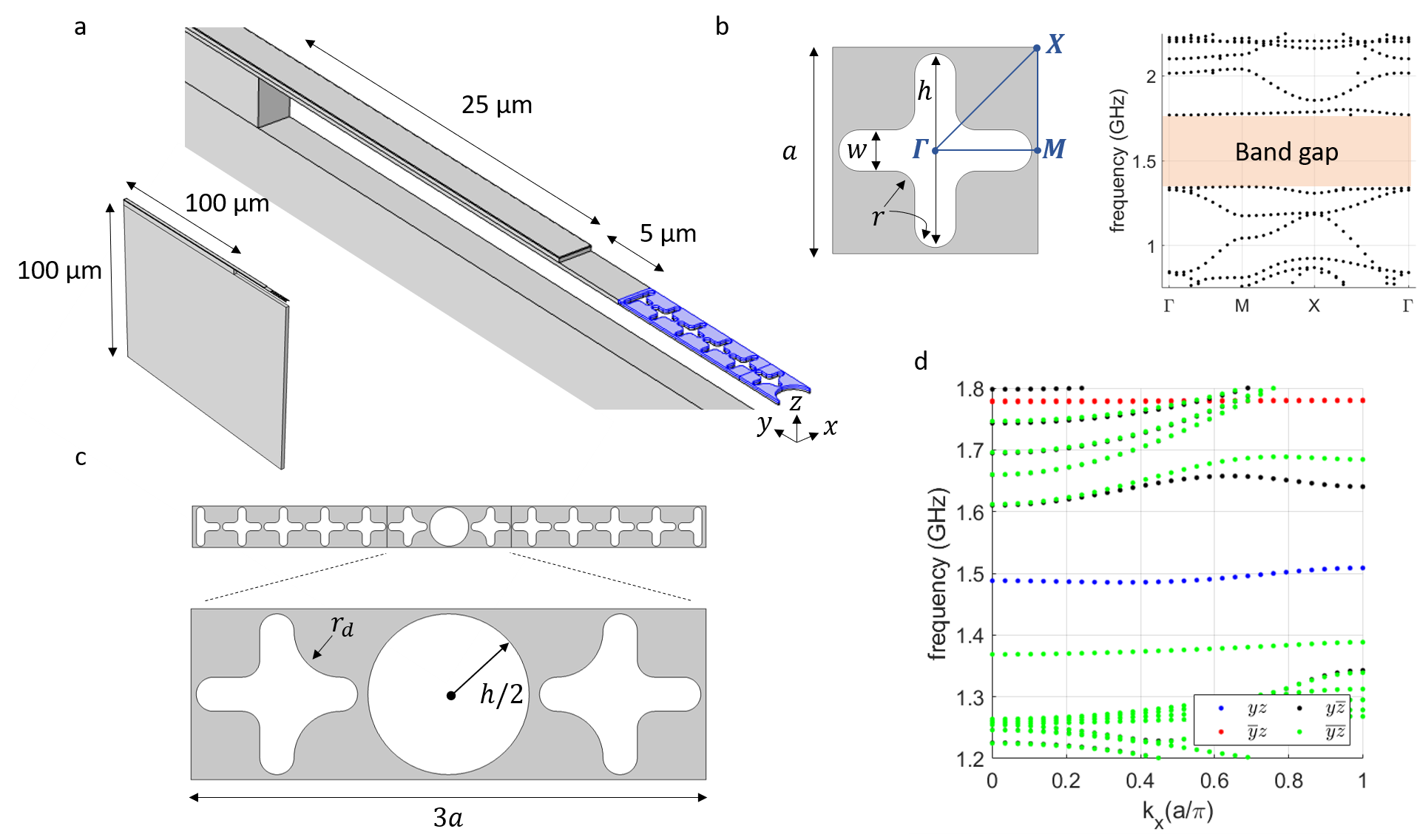}
    \caption{(a) Simulation geometry for the PnC waveguide system with bandgap surrounding 1.5 GHz. (b) Unit cell of the infinitely periodic 2D PnC and its bandstructure. (c) Plan view geometry of the PnC waveguide. (d) Waveguide bandstructure showing the symmetry properties of the modes. The overbars in the legend denote inversion symmetry for an origin at the geometrical center of the waveguide. The mode crossing 1.5 GHz (blue dots) is symmetric with respect to the $y$- and $z$-axes.}
    \label{fig:1.5GHz_structure}
\end{figure}

In contrast to the higher frequency system, the guided mode does not have increasing dispersion throughout the entire Brillouin zone [\figureautorefname{ \ref{fig:1.5GHz_guided_band}}]. The frequencies in the first half the Brillouin zone have degeneracy. These frequencies were not investigated for their phase shifting properties. Instead, we focused on the higher $k_x$ values that are nondegenerate. Displacement profiles of the guided mode at three values of $k_x$ are shown in \figureautorefname{ \ref{fig:1.5GHz_guided_band}}. The displacement is confined to the small linkages between neighboring circular holes, and emphasize that the band is constituted by a `tight-binding' type displacement field, leading to the relatively flat dispersion. 

\begin{figure}
    \centering
    \includegraphics[width=\textwidth]{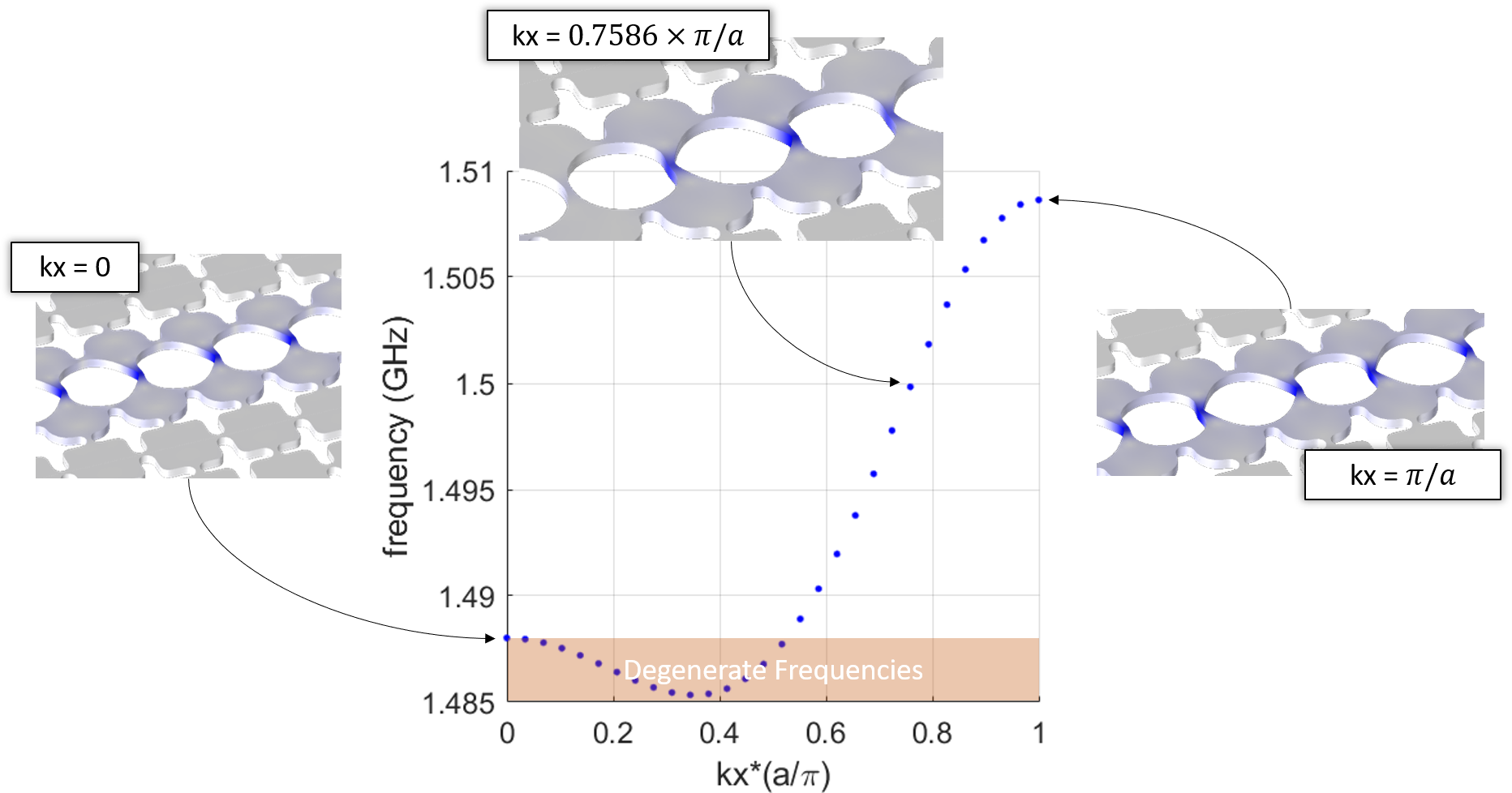}
    \caption{Plot of the guided band, including images of the displacement profile at 3 points along the band. Unlike the system described in the main text, this system includes a region with degenerate frequencies.}
    \label{fig:1.5GHz_guided_band}
\end{figure}

The performance of the 1.5 GHz waveguide system is shown in \figureautorefname{ \ref{fig:1.5GHz_results}}. Two aspects of this system stand out for consideration relative to the 5 GHz system: (1) the increased overhang of the piezoelectric actuator and the associated size of the suspended membrane structure and (2) the contrasting influence of the acoustoelastic effect on the lower frequency waveguide structure with larger structural features. The transfer function in \figureautorefname{ \ref{fig:1.5GHz_results}}(a) illustrates that the larger size of the suspended membrane results in a slower switching time, with the first resonance falling below 1 MHz; the nominal switching time for this system is about 1.5 \textmu s.  In \figureautorefname{ \ref{fig:1.5GHz_results}}(b), the frequency shift induced at $-20$ V is plotted along with the group velocity. Relative to the 5 GHz structure presented in the main text, the sign of the frequency shift at negative bias has switched. In addition, the $-20$ V bias level required to achieve $\Delta f$ of $\sim 5$ MHz is much lower than the $-50$ V needed for the 5 GHz structure. This increased response at negative bias is achieved due to the much longer overhanging span of the piezoelectric actuator, which extends for 25 \textmu m compared with just 5 \textmu m for the higher frequency device structure. To investigate $\Delta f$ as a function of bias, we focus on a spatial frequency $k_x = 0.7586\pi/a$ that has an operating frequency very close to 1.5 GHz and a group velocity $v_g = 243$ m/s that is near the maximum group velocity for the band. Sweeping the bias at this operating point, as shown in \figureautorefname{ \ref{fig:1.5GHz_results}(c)}, reveals that the improved tensile response comes at the expense of the performance in compression at positive bias. The bowing in the membrane under compression is so severe that there is only a limited range between 0 and +10 V where a positive phase shift (negative frequency shift) can be achieved. Thus, even though the length required for a $-\pi$ phase shift is less than 12 PnC periods long ($\sim 24$ \textmu m) at $-20$ V, the structure yields a limited capacity for positive phase shift at positive bias, demonstrating the tradeoff posed by the size of the overhanging membrane when engineering this class of devices. If only one implication of phase shifting is needed and the switching time is not important, increasing the size of the overhanging actuator will optimize performance. Finally, note that the acoustoelastic effect causes a much smaller offset in $\Delta f$ here relative to the 5 GHz system. When $W_3$ is included, an offset in $\Delta f$ that counteracts the second order term is obtained, just as for the 5 GHz system. However, the magnitude of the correction to $\Delta f$ is substantially smaller. In the 5 GHz system, the offset is so large that the sign of $\Delta f$ is switched, corresponding to a change in the implication of the phase shift. 

\begin{figure}
    \centering
    \includegraphics[width=\textwidth]{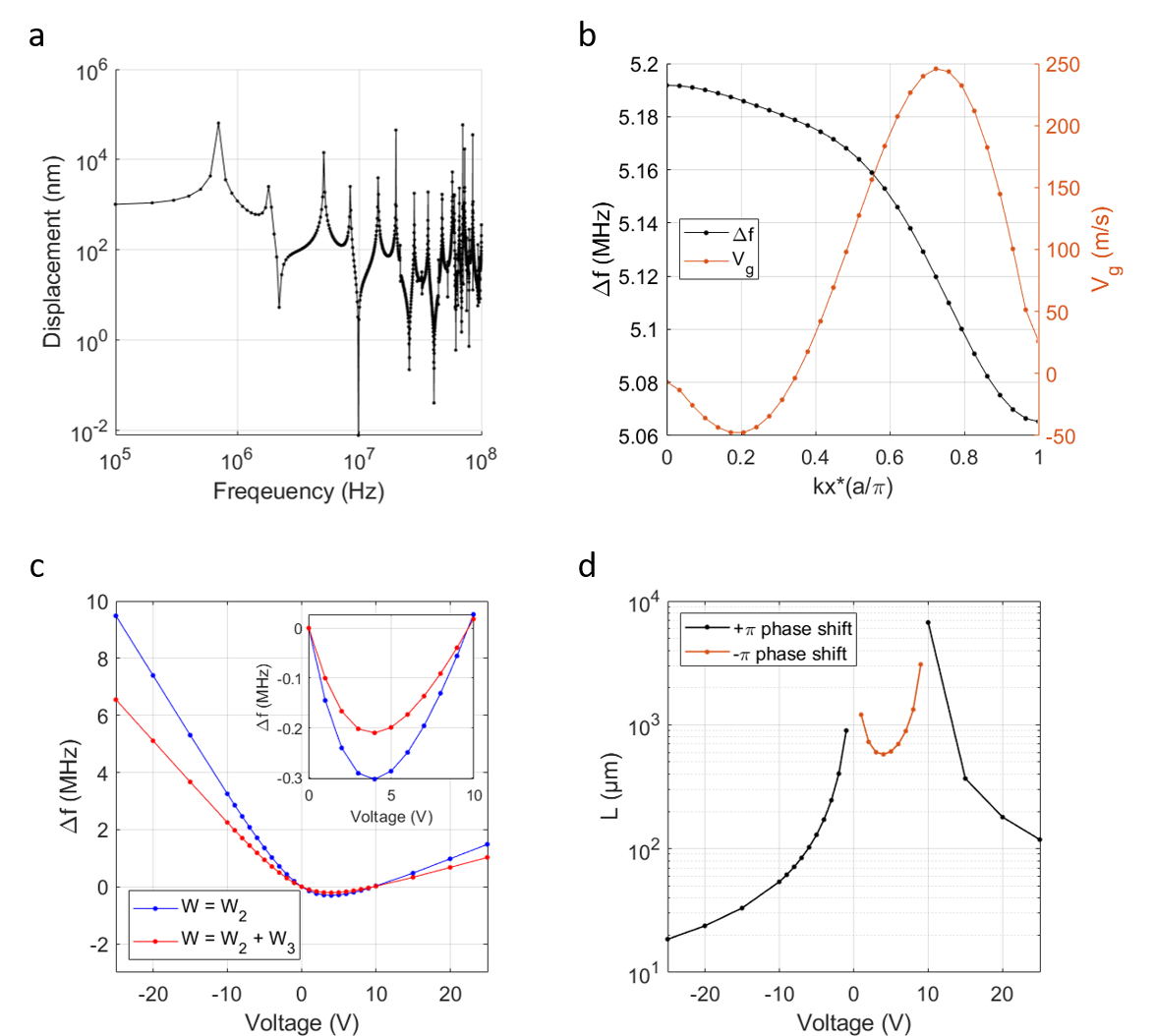}
    \caption{Simulation results for the 1.5 GHz waveguide system. (a) Transfer function for a bias of $20$ V, showing a marked increase in switching time. The first resonance lies at just above 700 kHz in frequency, with an associated switching time of nearly 1.5 \textmu s. (b) Frequency shift $\Delta f$ (black) induced by $-20$ V piezo bias and group velocity $v_g$ (orange) for the guided band. (c) Computed frequency shifts over a range of piezo biases for the two constitutive relations. Here, a propagation constant of $k_x = 0.7586\pi/a$ was used, corresponding to a zero bias temporal frequency of 1.4998 GHz. At this fixed spatial frequency, (d) shows the length of waveguide required for a $\pi$ phase shift when $W=W_2 + W_3$.}
    \label{fig:1.5GHz_results}
\end{figure}

\section{Calculation of the contribution to the phase shift caused by the change in path length $k\cdot \Delta L$}

To assess the purely mechanical phase shift arising from the change in path length, we construct a finite structure for simulation that models a 20-period phase shifter with input and output waveguides each extending for 100 periods. This system is shown in \figureautorefname{ \ref{fig:kDeltaL}}(a,b) at a piezo bias of $-50$ V, where the displacement has been scaled up by 10x. Note that the waveguide structure implemented here is for the 5 GHz phase shifter described in the main text, not the 1.5 GHz structure presented in the previous section. The $x$-component of the displacement field is plotted according to the color scale in \figureautorefname{ \ref{fig:kDeltaL}}(a) and the $z$-component is similarly plotted in \figureautorefname{ \ref{fig:kDeltaL}}(b). The displacement field $\mathbf{u} = (u,v,w)$ is the vector that relates the material frame $\mathbf{X}=(X,Y,Z)$ to the spatial frame $\mathbf{x}=(x,y,z)$ according to the relation $\mathbf{x}=\mathbf{X}+\mathbf{u}$. The symmetry-folded structure used in the computation is shown in \figureautorefname{ \ref{fig:kDeltaL}}(c), where the piezoelectric actuator extends for 10 periods and an additional length of 100 periods of waveguide follows. Accounting for the mirror planes yields the full structure, which is $y$-symmetric. Thus $v$ is identically 0 in the geometrical center of the waveguide where we measure the mechanical dilatation by extracting the displacement vector once each period, but the out-of-plane bulging of the membrane results in $w$ rising to a maximum that is 2 orders larger than that of $u$, as shown in \figureautorefname{ \ref{fig:kDeltaL}}(d). The green boxes in plots (d)--(f) of \figureautorefname{ \ref{fig:kDeltaL}} designate the 10 periods where the piezoelectric actuator is present in the reduced computational structure, meaning that the origin in these plots is in the middle of the phase shifter. The effect of $u$ on the change in path length is dominant in the vicinity of the piezoelectric actuators, but where the bowed structure elides down to the relatively unstrained waveguide far from the piezos, the large magnitude of $w$ results in a change of sign of $\Delta l$, the dilatation in path length per period.
\begin{equation}
    \Delta l = \sqrt{(\Delta u + a)^2 + (\Delta w)^2} - a
\end{equation}
The curves plotting the mechanical phase shift $k \cdot \Delta L$ in \figureautorefname{ \ref{fig:kDeltaL}}(e,f) are simply the integral of $\Delta l$ multiplied by different scaling factors. The phase shift caused by the path length dilatation peaks adjacent to the bulge in the membrane, but drops farther from the piezoelectric actuators and changes sign for the $-50$ V case. Thus we find that for our operating point of $k=0.6897 \cdot \pi/a$ and voltage of $-50$ V, the 20 period phase shifter yields a phase shift due to the mechanical change in path length $k \cdot \Delta L$ that is approximately  $2 \times 0.05^\circ = 0.1^\circ$, where the factor of 2 is included to account for the symmetry plane that doubles the length of the waveguide. By contrast, $\Delta k \cdot L \approx 97^\circ$ when the result for the periodic system is extrapolated to 20 waveguide periods.
\begin{figure}[ht]
    \centering
    \includegraphics[width=\textwidth]{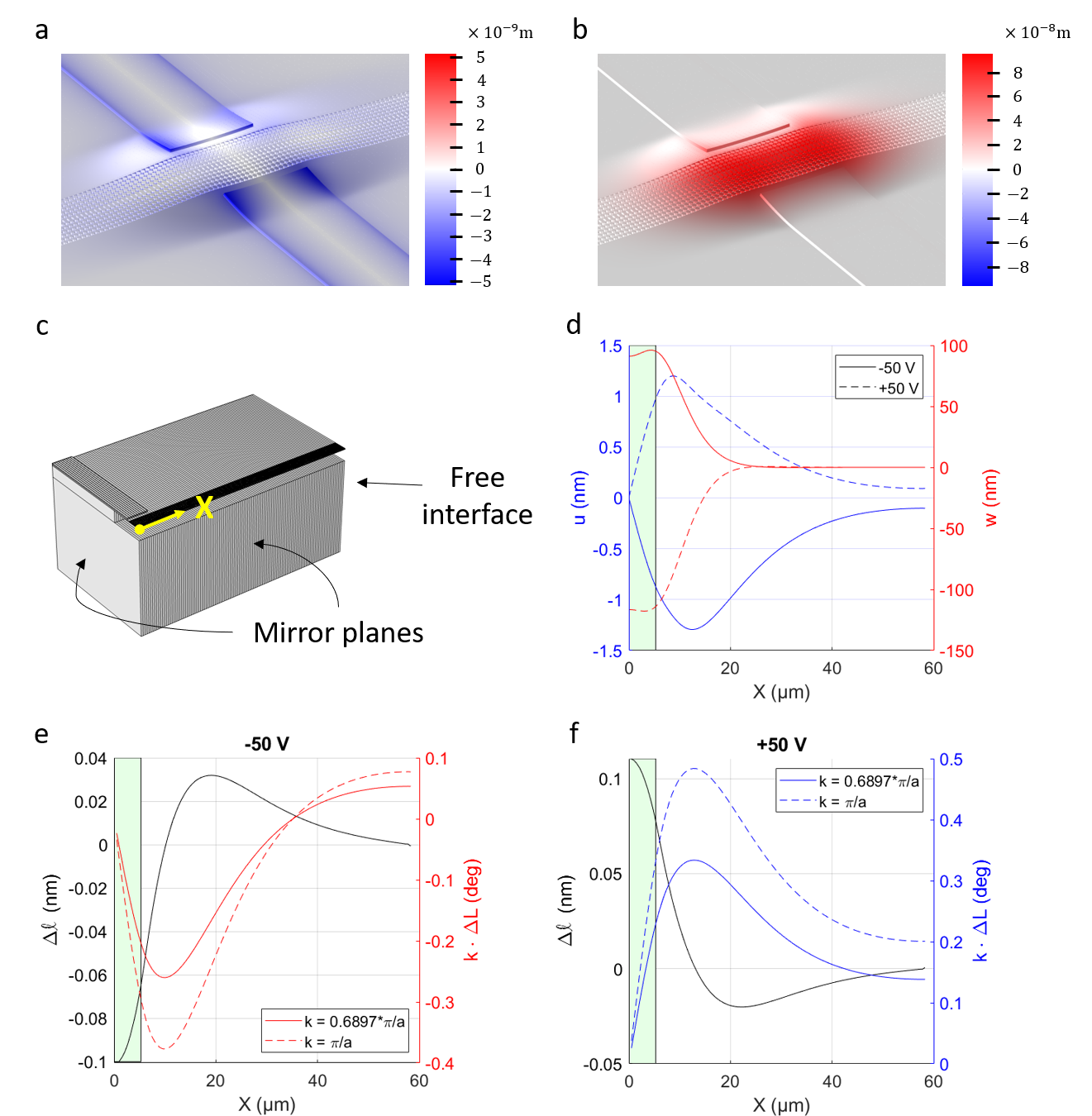}
    \caption{(a,b) Twenty period phase shifter under a bias of $-50$ V plotted with a 10x augmentation in the deformation, where $u$ is plotted according to the color scheme in (a) and $w$ is plotted similarly in (b). (c) The computational structure used, showing the symmetry planes that reduce the size of the structure. The $X$-axis designated by the yellow arrow is used for the abscissas of plots (d--f), which also include green boxes where the piezoelectric actuators surround the waveguide. (d) Plots of $u$ and $w$ for both $-50$ V and $+50$ V biases. (e,f) Analyses of the phase shift $k \cdot \Delta L$ induced by the path length dilatation for biases of $-50$ V and $+50$ V. These plots correspond to the half-structure. }
    \label{fig:kDeltaL}
\end{figure}

\section{Cavity design for dynamic coupling rate modification for quantum memory applications}

\figureautorefname{ \ref{fig:cavity}} shows a cavity which could be used to build the phononic memory described in the main text. The modal frequencies identified in \figureautorefname{ \ref{fig:cavity}}(a) are consistent with the bandgap of the 2D PnC, as shown in Figure 2 in the main text. The modes are designated by their inversion and reflection symmetry properties. This cavity spans a $3\times 3$ PnC structure in space and yields a mode that is isolated in frequency and shares the same symmetry properties as the waveguide mode [\figureautorefname{ \ref{fig:cavity}}(a)]. The frequency of the mode can be tuned by modulating the size of the major axis of the elliptical holes that are in the side-centered positions of the defect [\figureautorefname{ \ref{fig:cavity}}(c)]. The circular and elliptical holes in this structure are limited to angular sections modifying the cross geometries at lattice sites of the PnC adjacent to the position of a missing cross. The computational structure was defined by the defect geometry shown in \figureautorefname{ \ref{fig:cavity}}(c), with an additional border of 6 PnC unit cells. The outer border of the PnC structure was held fixed. The eigenfrequencies were computed with symmetry folding along the $x$-, $y$-, and $z$-axes, as shown in \figureautorefname{ \ref{fig:cavity}}(d).

\begin{figure}
    \centering
    \includegraphics[width=\textwidth]{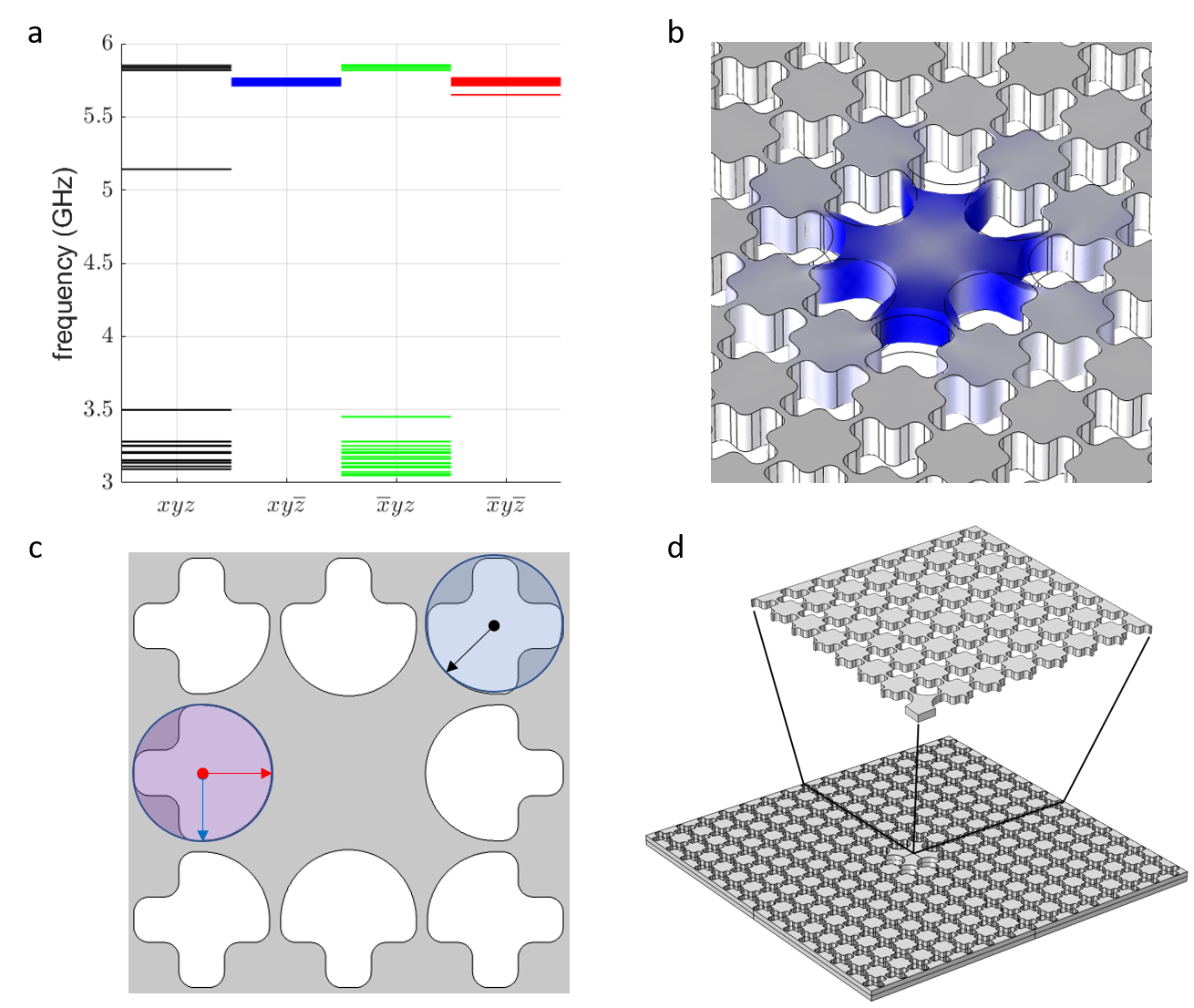}
    \caption{Simulation results and geometry for a phononic cavity at 5.14 GHz. (a) The modal spectrum for the system, partitioned according to symmetry constraints. The first column in black shows that there is a fully symmetric mode at 5.14 GHz. The overbars in the labels along the ordinate axis denote inversion symmetry, while the bare coordinate labels represent reflection symmetry. There are no other modes near the 5.14 GHz mode within the bandgap of the PnC. (b) Displacement profile of the symmetric mode at 5.14 GHz. (c) Plan view geometrical definition of the defect region. A 90 degree sector of a circle with radius $h/2$ was used to define the defects in the corner positions, while a 180 degree sector of an ellipse with minor axis (blue) of $h/2$ and major axis (red) of $1.03h/2$ was used to define the centered positions. (d) The computations were accomplished with symmetry folding, yielding the reduced structure shown in the call out.}
    \label{fig:cavity}
\end{figure}

\section{SLH derivation}

Consider the diagram shown in Figure~\ref{fig:scheme}.
\begin{figure}[!tb]
	\centering
	\includegraphics[width=0.7\textwidth]{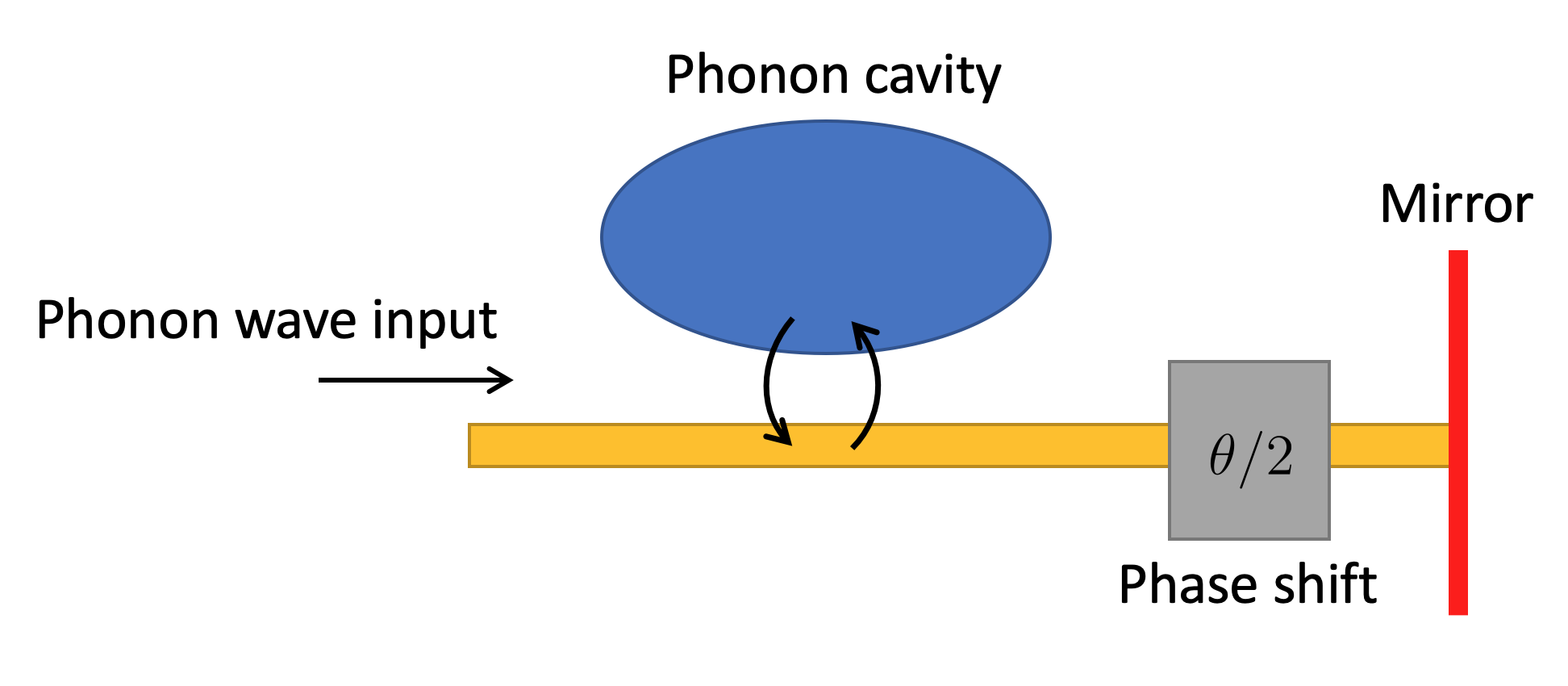}
	\caption{Coupling scheme between a flying incoming qubit (phonon wave) and a stationary qubit supported by a phonon cavity via a phase-inducing feedback element.} \label{fig:scheme}
\end{figure}
A flying qubit in a phonon waveguide interacts with a phonon cavity. The cavity has two input-output ports. The left and right portions of waveguide both constitute inputs to and outputs from the cavity. The right-ward moving output leaving the cavity travels into the phase shifter where it undergoes a single-pass phase shift of $\theta/2$ before reflecting from a PnC mirror and passing through the phase shifter another time. Hence, the total accumulated phase is $\theta$. Finally, the phonon wave propagates to the left in the waveguide, interacting with the phonon cavity one more time. The overall system output is the wave flying towards the left in the waveguide after leaving the left input-output interface of the cavity. To analyze the dynamics of this tunable resonant cavity, we will use the SLH triplet parameter set~\cite{combes_slh_2017} to formulate a model that includes each element of the system and then use the SLH connection rule to derive and solve the dynamical equation for the system variables. This process enables us to analyze the achievable quantum state fidelity of the system. 

There are two input-output subsystems in \figureautorefname{ \ref{fig:scheme}}: the cavity and the phase shifter. The cavity has two input-output ports and a single internal degree of freedom that is quantized by the cavity action. Let us denote the fixed input-output coupling rate between the waveguide and the cavity as $\kappa_e$ and the coupling between the cavity's intrinsic loss port and the environment as $\kappa_i$. The phase shifting element and the mirror together have a single input-output interface. Combined, they constitute a simple phase-shifting element with an induced phase $\theta$. We assume that the intrinsic loss in this subsystem is negligible.

To establish the SLH triplets for the phonon cavity and phase shifter, we define the naming of the input and the output interfaces as follows. The wave moving into the cavity from the left is `input 1' and the wave moving out of the cavity towards the right is `output 1'. Similarly, the wave moving into the cavity from the right is `input 2' and the wave moving out of the cavity towards the left is `output 2'.  The intrinsic loss is designated by `input 3' and `output 3'. Then, the SLH-triplet representing the phonon cavity is~\cite{combes_slh_2017}
\begin{equation}
	\vt{G}_1 = \left( \vt{S}_1, \vt{L}_1, H_1 \right) = \left( \left( \begin{array}{ccc} 1 & 0 & 0 \\ 0 & 1 & 0 \\ 0 & 0 & 1  \end{array} \right), \left( \begin{array}{c} \sqrt{\kappa_e} a_c \\ \sqrt{\kappa_e} a_e \\ \sqrt{\kappa_i} a_c  \end{array} \right), \hbar \omega_c a^\dagger_c a_c \right),
\end{equation}
where $\hbar$ is the reduced Planck constant, $\omega_c$ is the resonant frequency of the phonon cavity, and $a_c$ is the annihilation operator for the bosonic quantized field in the phononic cavity, satisfying the commutation rule $[a_c, a^\dagger_c] = 1$. The SLH-triplet representing the phase shifter is~\cite{combes_slh_2017}
\begin{equation}
	\vt{G}_2 = (\vt{S}_2 , \vt{L}_2 , H_2) = (e^{i \theta}, 0, 0),
\end{equation}
which has only a single input-output port. 

To connect these subsystems following the structure shown in \figureautorefname{~\ref{fig:scheme}}, we first start with $\vt{G}_1$ and $\vt{G}_2$. Note, however, that $\vt{G}_1$ has three input-output ports while $\vt{G}_2$ has only a single input-output port. To address this discrepancy, we make an auxiliary trivial system in parallel to $\vt{G}_2$,
\begin{equation}
	\vt{2} = \left( \left( \begin{array}{cc} 1 & 0 \\ 0 & 1 \end{array} \right), \left( \begin{array}{c} 0 \\ 0 \end{array} \right), 0 \right).
\end{equation}
This is a trivial two input-output port system with no internal degree of freedom or loss ports. Then, the connection is modeled as
\begin{equation}
	(\vt{G}_2 \boxplus \vt{2}) \triangleleft \vt{G}_1.
\end{equation}
The above connection rule means that output 1 from $\vt{G}_1$ is connected to the input of $\vt{G}_2$ while the other two outputs from $\vt{G}_1$ are connected to the trivial system $\vt{2}$.  Next, we connect the first output of the above connected system to the input 2 of $\vt{G}_1$ since the output of the phase shifter is directly connected to the right input to the cavity. This geometry of connection is accomplished by
\begin{equation}
	\vt{G}_t = \left( (\vt{G}_2 \boxplus \vt{2}) \triangleleft \vt{G}_1 \right)_{[1 \rightarrow 2]}. 
\end{equation}
The block diagram of this connected system is shown in \figureautorefname{~\ref{fig:block}}. 

\begin{figure}[!tb]
	\centering
	\includegraphics[width=0.8\textwidth]{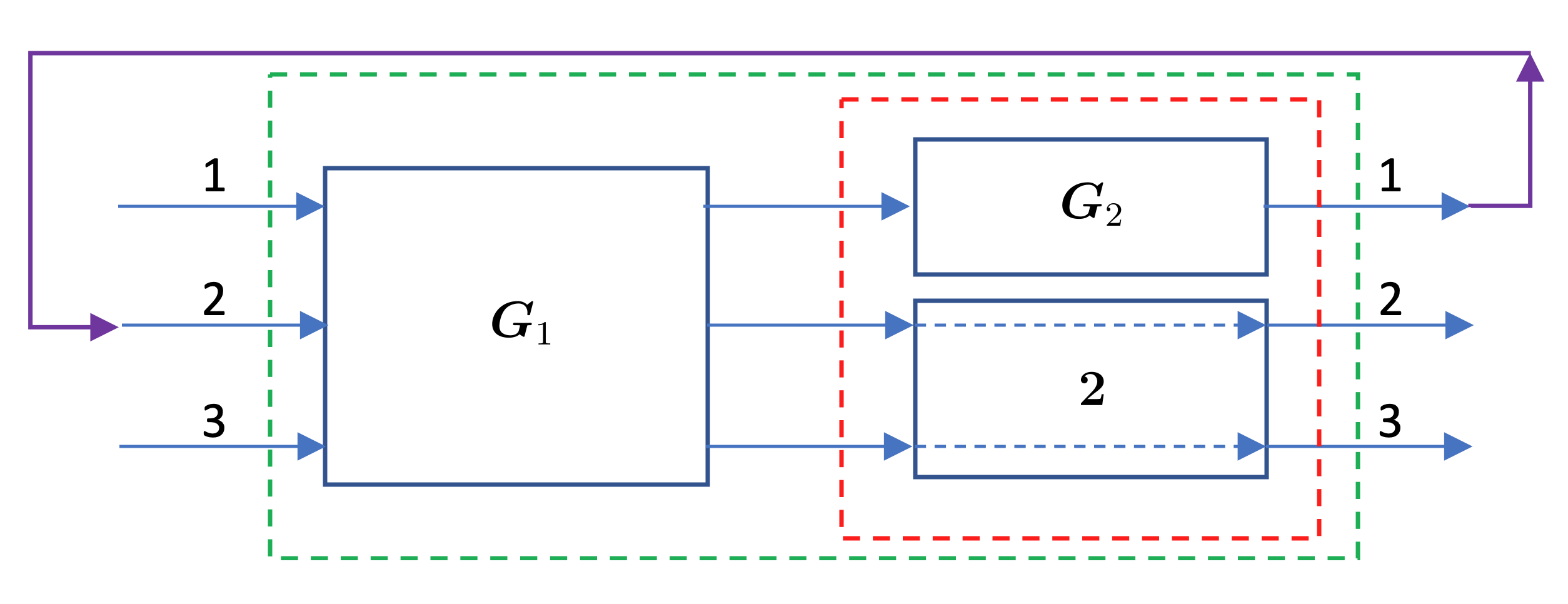}
	\caption{Block diagram of system connection. The new system inside the dashed red box corresponds to $(\vt{G}_2 + \vt{2})$. The feedback connection is from the first output of the combined system $((\vt{G}_2 + \vt{2}) \triangleleft \vt{G}_1 )$ (inside the green dashed box) to the second input of the same combined system.} \label{fig:block}
\end{figure}

Applying the reduction rule, we obtain the reduced SLH-triplet for the entire system on the rotating frame (i.e., $\omega_c \rightarrow 0$)
\begin{equation}
	\vt{G}_t = (\vt{S}_t, \vt{L}_t, H_t)  = \left( \left( \begin{array}{cc} e^{i \theta} & 0 \\ 0 & 1 \end{array} \right) , \left( \begin{array}{c} \sqrt{\kappa_e} (e^{i \theta} + 1) a_c \\ \sqrt{\kappa_i} a_c \end{array} \right) , \kappa_e \sin \theta a_c^\dagger a_c \right),
\end{equation}
where the first and second basis elements represent the net cavity input-output and the intrinsic loss channel, respectively. Recall that the scattering matrix describes the transformation of the input to the output in the absence of cavity population, while the Lindbladian describes the output in the absence of an input signal. Intuitively, given vacuum input, the first element of the composite Lindbladian comports with input-output theory, since the first term $\sqrt{\kappa_e} e^{i \theta} a_c$ represents the leftward-traveling input (equivalent to the rightward-traveling output $\sqrt{\kappa_e} a_c$ after a phase shift $\theta$), while the second term $\sqrt{\kappa_e} a_c$ represents the raw leftward traveling output rate. The interference between these two signals generates the final composite output through the left port. On the other hand, given an empty cavity, the $\theta$ phase shift from the initial input to the final output as shown in the scattering matrix is explained by the fact that the signal undergoes such a shift during the cavity-mirror round trip. 

The feedback system now has a $\theta$-dependent Hamiltonian $H_t = \kappa_e \sin \theta a^\dagger_c a_c$, and the operator differential equation in the Heisenberg picture for $a_c$~\cite{combes_slh_2017} yields
\begin{equation}
	\dot{a}_c = - i \kappa_e \left( \sin \theta \right) a_c - \kappa_e \left( 1 + \cos \theta \right) a_c - \frac{1}{2} \kappa_i a_c - \sqrt{\kappa_e} \left( e^{i \theta} + 1 \right) a_\mathrm{in} - \sqrt{\kappa_i} f_i.
\end{equation}
The first term on the righthand side of this equation is the phase rotational component of the Hamiltonian, which has a frequency detuning of $\kappa_e \sin \theta$. The second term represents the overall decay rate through the main output, which propagates in the waveguide towards the left from the cavity at a rate of $2 \kappa_e (1 + \cos \theta)$. The third term is the intrinsic loss at the rate $\kappa_i$, and the fourth term involves the complex input coupling rate $\sqrt{\kappa_e} (e^{i \theta} + 1)$ and the input field $a_\mathrm{in}$. Note that this input coupling rate implies that the input \textit{power} coupling rate is $|\sqrt{\kappa_e} (e^{i \theta} + 1)|^2 = 2 \kappa_e (1 + \cos \theta)$, which matches the decoherence rate of $a_c$ due to the input-output coupling. The last term involves the Wiener process (white noise) $f_i$ and its amplitude coupling rate $\sqrt{\kappa_i}$. This is the intrinsic loss channel. Moreover, the input-output relation is~\cite{combes_slh_2017}
\begin{equation}
	a_\mathrm{out}(t) = e^{i \theta} a_\mathrm{in} (t) + \sqrt{\kappa_e} (e^{i \theta} + 1) a_c (t).
\end{equation}
Physically, the output wave $a_\mathrm{out}$ can be conceptualized as resulting from an interference of 3 waves at the left output port of the cavity: the input wave $a_\mathrm{in}$ (phase-shifted by $\theta$ due to the round trip to the mirror and back), the raw rightward-traveling output $\sqrt{\kappa_e} a_c$ (also phase shifted by $\theta$ due to the round trip), and the raw leftward-traveling output $\sqrt{\kappa_e} a_c$ (with no phase shift). The feedback from the phase shifter to the cavity provides adjustable input coupling, which is represented as a Lindblad operator:
\begin{equation}
	\sqrt{\kappa_e} \left( e^{i \theta} + 1 \right) a_c = 2 e^{i \theta/2} \sqrt{\kappa_e} \cos (\theta/2) a_c. 
\end{equation}
The effective total raw output \textit{power} rate $\kappa_\mathrm{eff}$ is the amplitude-squared of the raw output amplitude rate:
\begin{align}
\kappa_\mathrm{eff} &= \Big|2 e^{i\theta/2} \sqrt{\kappa_e} \cos{(\theta/2)}\Big|^2 \\
&= 2 \kappa_e (1 + \cos{\theta}).
\end{align}
Note that the maximum raw output power rate is $4\kappa_e$ rather than $\kappa_e$. This is due to the fact that $\kappa_e$ represents the raw output rate of the cavity through \textit{each} port, and that the raw rightward-traveling output is eventually reflected by the phase-shifting mirror and returns to interfere with the raw leftward-traveling output of the cavity. If the two waves constructively interfere, then the amplitude doubles, thus quadrupling the power. Provided that $\theta$ is adjustable in time, the raw output rate can be completely controlled between zero and $4 \kappa_e$ as $\theta$ varies between $\pi$ and $0$, respectively.

There are two undesirable consequences of this feedback scheme which can be addressed by modifying the system. First, the input signal acquires an extra phase factor $e^{i \theta/2}$ relative to the net raw cavity output. Second, the final connected system takes on a detuning frequency of $\kappa_e \sin \theta$. Before we resolve these issues, we note that the controllable phase $\theta(t)$ can be chosen according to a recipe that optimizes the quantum state transfer from the flying input qubit to the stationary cavity qubit. Let us assume that the optimal profile $\theta(t)$ is known. 

We first tackle the problem posed by the unwanted extra phase of $e^{i \theta (t) /2}$ in the input coupling. This extra phase can be cancelled out by including an extra phase shifter that operates in reverse to induce a phase shift of $e^{-i \theta(t)/2}$ in front of input 1. Thus, achieving this reversed phase shift necessitates having a phase shifting device that can operate in `compression', as described previously. The additional phase shifter in the input waveguide can be represented by adding two additional phase shifting elements to the scheme shown in \figureautorefname{~\ref{fig:block}}. First,
\begin{equation}
	\vt{G}_0 = (\vt{S}_0 , \vt{L}_0, H_0) = (e^{-i \theta (t)/2}, 0,0 )
\end{equation}
should be placed in front of the input 1 of the system. Second, 
\begin{equation}
	\vt{G}_3 = (\vt{S}_3, \vt{L}_3, H_4) = (e^{-i \theta(t)/2}, 0,0),
\end{equation}
must follow after the output 2 of the system, thus bringing the output signal in-phase with the raw cavity output and the previously phase-shifted input signal.

Next, we address the undesired detuning frequency $\kappa_e \sin \theta(t)$. To null out this detuning, the cavity can be detuned in the opposite direction. This is best represented by a time dependent Hamiltonian of the phonon cavity, such that the new SLH-triplet for the modified phonon cavity is now
\begin{equation}
	\vt{G}'_1 = (\vt{S}'_1, \vt{L}'_1, H'_1) =  \left( \left( \begin{array}{ccc} 1 & 0 & 0 \\ 0 & 1 & 0 \\ 0 & 0 & 1  \end{array} \right), \left( \begin{array}{c} \sqrt{\kappa_e} a_c \\ \sqrt{\kappa_e} a_e \\ \sqrt{\kappa_i} a_c  \end{array} \right), - \kappa_e \sin(\theta(t)) a^\dagger_c a_c \right),
\end{equation}
on a rotating frame (i.e., $\omega_c \rightarrow 0$). Then, the new system is constructed through the following recipe:
\begin{equation}
	\vt{G}'_t = \left( \left( \vt{1} \boxplus \vt{G}_3 \boxplus \vt{1}  \right)\triangleleft \left( \left( \vt{G}_2 \boxplus \vt{2} \right)\triangleleft \left(  \vt{G}'_1 \triangleleft \left( \vt{G}_0 \boxplus \vt{2} \right)\right) \right) \right)_{[2 \rightarrow 1]}.
\end{equation}
Figure~\ref{fig:block2} shows a block diagram representation of the above system. 

\begin{figure}
	\centering
	\includegraphics[width=\textwidth]{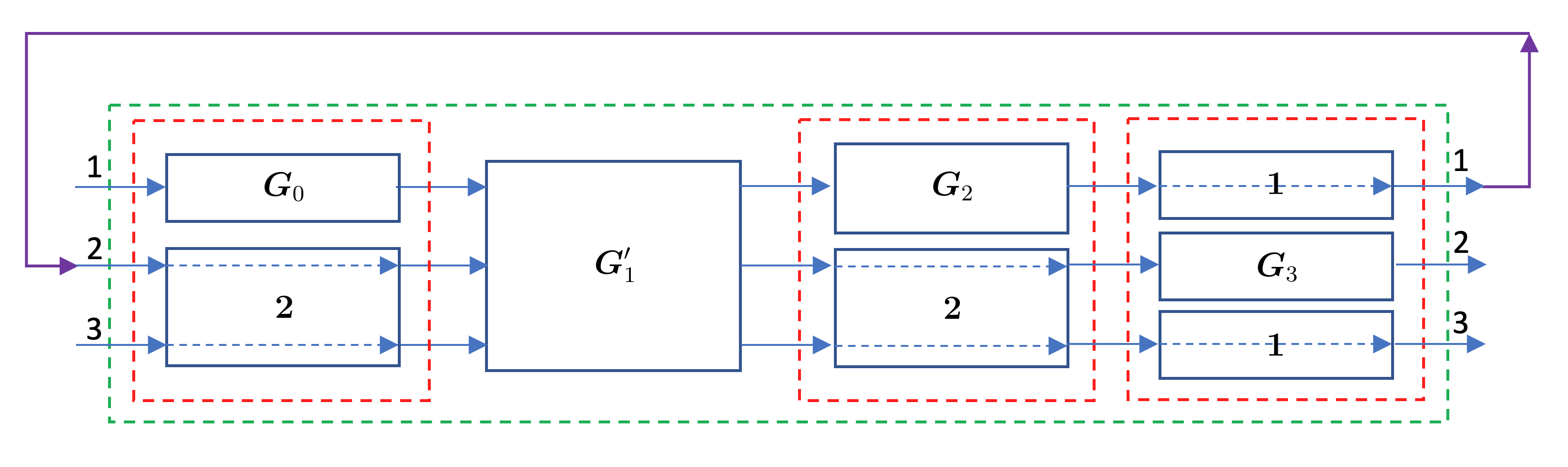}
	\caption{Modified block diagram including additional phase shifter at the input port to the cavity.} \label{fig:block2}
\end{figure}

Finally, we apply the SLH connection rule to the above equation and obtain the entire system's SLH triplet:
\begin{equation}
	\vt{G}'_t = (\vt{S}_t, \vt{L}_t, H_t)  = \left( \left( \begin{array}{cc} 1 & 0 \\ 0 & 1 \end{array} \right) , \left( \begin{array}{c} 2 \sqrt{\kappa_e} \cos (\theta/2) a_c \\ \sqrt{\kappa_i} a_c \end{array} \right) , 0 \right).
\end{equation}
The corresponding dynamic equation is given as
\begin{equation}
	\dot{a}_c = - \frac{1}{2} \left( 4 \kappa_e \cos^2 \left( \frac{\theta(t)}{2} \right) + \kappa_i \right) a_c - 2 \sqrt{\kappa_e} \cos \left( \frac{\theta(t)}{2} \right) a_\mathrm{in} - \sqrt{\kappa_i} f_i.
\end{equation}
Furthermore, the input-output relation becomes
\begin{equation}
	a_\mathrm{out} (t) = a_\mathrm{in} (t) + 2 \sqrt{\kappa_e} \cos \left( \frac{\theta(t)}{2} \right) a_c.
\end{equation}
We have now suppressed the side effects of the feedback scheme arising from the system shown in Figure~\ref{fig:scheme} by correcting the detuning frequency and extra phase accumulation, leading to more conventional resonant cavity quantum dynamics. The modified system is diagrammed in Figure 5(b) in the main text. Here, the output amplitude coupling rate is given as $2\sqrt{\kappa_e} \cos (\theta(t)/2)$, which is a real number, and the corresponding power coupling rate is given as $4 \kappa_e \cos^2 (\theta(t)/2) = 2 \kappa_e ( 1 + \cos (\theta(t)))$.

\section{Optimizing the Phase Profile}
\label{sec: Optimizing the Phase Profile}

We now aim to optimize the temporal profile of the phase $\theta(t)$, which is controllable by tuning the distance between the phonon cavity and the mirror. To do so, we recall our previous analysis regarding the ideal temporal profile of the output coupling rate for a resonator~\cite{chatterjee_optimal_2020} in order to maximally absorb an input signal. Given an exponential input profile $r_{in}(t) = re^{-rt}$, a maximum physically achievable output coupling rate $\kappa_{max}$, and a resonator intrinsic loss rate $\kappa_i$, the optimal output coupling rate profile $\kappa(\tau)$ (where $\tau = \kappa_{max}t$ represents the time converted to a dimensionless value) is determined as follows~\cite{chatterjee_optimal_2020}:
\begin{align} \label{eq: kappa(t) exponential input}
\begin{split}
\kappa(\tau) =
\begin{cases}
\kappa_{max}, & 0 < \tau < \tau_c \\
\kappa_{max} \frac{r}{\kappa_{max}} \Bigg(A_1 e^{\Big(\frac{r}{\kappa_{max}} - \frac{\kappa_i}{\kappa_{max}}\Big) \tau} - A_2\Bigg)^{-1}, & \tau > \tau_c
\end{cases},
\end{split}
\end{align}
where the constants $A_1$ and $A_2$ are the following:
\begin{align}
A_1 &= \frac{4}{\Big(\sqrt{\frac{\kappa_{max}}{r}} - \sqrt{\frac{\kappa_i}{\kappa_{max}} \frac{\kappa_{max}}{r} \frac{\kappa_i}{\kappa_{max}}} + \sqrt{\frac{r}{\kappa_{max}}}\Big)^2} \Bigg(\frac{2}{1 - \frac{\kappa_i}{\kappa_{max}} + \frac{r}{\kappa_{max}}}\Bigg)^{-\frac{2}{1 + \frac{\kappa_i}{\kappa_{max}} - \frac{r}{\kappa_{max}}}} \\
&\quad + \frac{1}{1 - \frac{\kappa_i}{\kappa_{max}} \frac{\kappa_{max}}{r}} \Bigg(\frac{2}{1 - \frac{\kappa_i}{\kappa_{max}} + \frac{r}{\kappa_{max}}}\Bigg)^{\frac{2\Big(\frac{\kappa_i}{\kappa_{max}} - \frac{r}{\kappa_{max}}\Big)}{1 + \frac{\kappa_i}{\kappa_{max}} - \frac{r}{\kappa_{max}}}}, \\
A_2 &= \frac{1}{1 - \frac{\kappa_i}{\kappa_{max}} \frac{\kappa_{max}}{r}}.
\end{align}
and the threshold time $\tau_c$ is the following:
\begin{equation}
\tau_c = \frac{2}{1 + \frac{\kappa_i}{\kappa_{max}} - \frac{r}{\kappa_{max}}} \ln{\Bigg(\frac{2}{1 - \frac{\kappa_i}{\kappa_{max}} + \frac{r}{\kappa_{max}}}\Bigg)}.
\end{equation}
Note that the optimal output coupling profile normalized to the maximal physically achievable output coupling rate $\kappa_{max}$ varies specifically with the input rate and loss rates normalized to $\kappa_{max}$, rather than with those values themselves. As mentioned in the previous section, the output coupling rate $\kappa(\tau)$ relates to the phase $\theta(\tau)$ as follows:
\begin{equation}
\kappa(\tau) = 2 \kappa_e ( 1 + \cos (\theta(\tau))).
\end{equation}
Therefore, the maximum physically achievable rate $\kappa_{max}$ is attained when $\cos (\theta(\tau)) = 1$, yielding $\kappa_{max} = 4\kappa_e$, which we will substitute into the above expressions. We also note that the intrinsic loss rate $\kappa_i$ is negligible compared to $\kappa_e$ and the inverse rise time $r$. We thus convert our time units to the new dimensionless value $\tau' = \kappa_e t$ and re-write the optimal $\kappa$ as follows:
\begin{align} \label{eq: kappa(t) 2}
\begin{split}
\kappa(\tau') =
\begin{cases}
4\kappa_e, & 0 < \tau' < \tau'_c \\
\kappa_e \frac{r}{\kappa_e} e^{-\frac{r}{\kappa_e}\tau'} \Big(A_1 - A_2 e^{-\frac{r}{\kappa_e}\tau'}\Big)^{-1}, & \tau' > \tau'_c\end{cases},
\end{split}
\end{align}
where $A_1$ and $A_2$ reduce to the following:
\begin{align}
A_1 &= \frac{16}{\Big(4\sqrt{\frac{\kappa_e}{r}} + \sqrt{\frac{r}{\kappa_e}}\Big)^2} \bigg(\frac{8}{4 + \frac{r}{\kappa_e}}\bigg)^{-\frac{8}{4 - \frac{r}{\kappa_e}}} + \bigg(\frac{8}{4 + \frac{r}{\kappa_e}}\bigg)^{-\frac{2}{4\frac{\kappa_e}{r} - 1}}, \\
A_2 &= 1,
\end{align}
and $\tau'_c$ simplifies as follows:
\begin{equation}
\tau'_c = \frac{2}{4 - \frac{r}{\kappa_e}} \ln{\bigg(\frac{8}{4 + \frac{r}{\kappa_e}}\bigg)}.
\end{equation}
Due to the lack of intrinsic loss, nearly all of the overall loss occurs in the first stage ($t < t_c$) when the resonator is gaining an initial seed population while incurring non-zero loss through the input-output channel. As discussed in~\cite{chatterjee_optimal_2020}, the fidelity in this case is determined as $F = A_1$.
\par 
Next, we calculate $\theta(\tau')$ from $\kappa(\tau')$:
\begin{equation}
\theta(\tau') = \cos^{-1} \bigg(\frac{\kappa(\tau')}{2\kappa_e} - 1\bigg).
\end{equation}
Note that the phase is a function of the output coupling rate in units of the maximal rate. In substituting for $\kappa(\tau')$, the optimal phase profile becomes a function of the maximum input rate $r$ in units of the maximal output coupling rate:
\begin{align} \label{eq: theta(tau')}
\begin{split}
\theta(\tau') =
\begin{cases}
0, & 0 < \tau' < \tau'_c \\
\cos^{-1} \bigg(\frac{r}{2\kappa_e} e^{-\frac{r}{\kappa_e}\tau'} \Big(A_1 - A_2 e^{-\frac{r}{\kappa_e}\tau'}\Big)^{-1} - 1\bigg), & \tau' > \tau'_c
\end{cases}.
\end{split}
\end{align}
The fidelity thus becomes a function of $r/\kappa_e$, as depicted in Figure 6(a) in the main text. For the selected numerical values $\kappa_e = 2\pi \times 300 \textrm{ kHz}$ and $r = 2\pi \times 100 \textrm{ kHz}$, the fidelity is 96.9\%, and the optimal phase profile $\theta(t)$ takes the form shown in Figure 6(b) in the main text. The threshold time at which we switch from the lossy stage (where we generate sufficient initial population in the resonator such that the resonator output can thereafter cancel out the reflected input signal through destructive interference) to the zero-loss stage is $t_c = 0.18 \textrm{ }$ \textmu s.

\section{Simulating the Effect of Time Delay}

Here, we semiclassically simulate how a non-negligible round trip time from the rightward-traveling output to the mirror and back to the leftward-traveling input affects the fidelity for the overall quantum transfer. We label the signals by \textit{port} rather than propagation direction. Specifically, we use the labels $l_{in,c}$, $r_{out}$, $r_{in}$, and $l_{out,c}$ to represent the rightward-traveling input to the cavity (through the left port), the rightward-traveling output from the cavity (through the right port), the leftward-traveling input to the cavity (through the right port), and the leftward-traveling output from the cavity (through the left port). We start by applying first-principles input-output theory for the rightward-traveling and leftward-traveling waves, respectively:
\begin{equation} \label{eq: left-to-right input-output}
r_{out} = l_{in,c} + \sqrt{\kappa_e} a_c,
\end{equation}
\begin{equation} \label{eq: right-to-left input-output}
l_{out,c} = r_{in} + \sqrt{\kappa_e} a_c.
\end{equation}
Note that the rightward-traveling input has passed through a phase shifter prior to reaching the cavity, while the leftward-traveling output passes through the same phase shifter after leaving the cavity. As such, $l_{in,c}$ and $l_{out,c}$ relate to the overall input $l_{in}$ and output $l_{out}$ in the following manner:
\begin{equation} \label{eq: left input phase shifter}
l_{in,c} = e^{-i\frac{\theta_1}{2}} l_{in},
\end{equation}
\begin{equation}
l_{out} = e^{-i\frac{\theta_1}{2}} l_{out,c},
\end{equation}
where $-\theta_1/2$ represents the phase shift generated by the phase shifter at any given time. By default, $\theta_1(t) = \theta(t)$, where $\theta(t)$ denotes the phase shift from the rightward-traveling output to the leftward-traveling input due to the round trip to the mirror and back.
\par 
Next, we consider the Hamiltonian for the resonator. If there were no detuning between the cavity field and the traveling fields, this would be zero in the rotating frame. However, since we have introduced dynamic detuning to the cavity such that the detuning at time $t$ is $-\kappa_e \sin{(\theta_2(t))}$, the Hamiltonian becomes the following:
\begin{equation}
H = -\hbar \kappa_e \sin{(\theta_2(t))} a_c^{\dag} a_c.
\end{equation}
Again, by default, $\theta_1(t) = \theta_2(t)$. We are now ready to apply the master equation given a set of input-output channels $n$, each with coupling rate $\kappa_n$:
\begin{align} \label{eq: dot A_c setup}
\begin{split}
\dot{A_c}(t) &= \expect{\dot{a_c}} \\
&= \expect{-\frac{i}{\hbar} [a_c,H] - \sum_n \bigg(\sqrt{\kappa_n} a_{in,n} + \frac{1}{2} \kappa a \bigg)} \\
&= \expect{i \kappa_e \sin{(\theta_2(t))} a_c - \sqrt{\kappa_e} r_{in} - \frac{1}{2} \kappa_e a_c - \sqrt{\kappa_e} l_{in,c} - \frac{1}{2} \kappa_e a_c} \\
&= i \kappa_e \sin{(\theta_2(t))} A_c(t) - \kappa_e A_c(t) - \sqrt{\kappa_e} R_{in}(t) - \sqrt{\kappa_e} e^{-i\frac{\theta_1(t)}{2}} L_{in}(t),
\end{split}
\end{align}
where we have substituted the relationship from Eq.~\eqref{eq: left input phase shifter} in the last line. Recall that our goal is to treat the entire system consisting of the phase shifter, cavity, and mirror as a single unit, with a single input-output channel on the left. As such, the leftward-traveling input $R_{in}$ represents an intermediate process which we wish to relate to the rightward-traveling input $L_{in}$ and the cavity population parameter $A_c$. To this end, we note that the leftward-traveling input simply represents the phase-shifted echo of the rightward-traveling output, with the former lagging the latter by the round-trip time $\delta_F$:
\begin{align} \label{eq: right output to right input}
R_{in}(t) &= e^{i\theta(t - \delta_F/2)} R_{out}(t-\delta_F) \\
&= e^{i\theta_1(t - \delta_F/2 - \delta_M)} R_{out}(t-\delta_F)
\end{align}
It is worth analyzing the time argument $t - \delta_F/2 - \delta_M$ for the phase acquired during the round trip. If $\theta_1(t) = \theta(t)$ (i.e. if the temporal phase profiles for the left phase shifter and the mirror are in tandem with one another), then the time argument for the acquired phase from the mirror should be $t - \delta_F/2$, since the signal takes a half-round-trip time $\delta_F/2$ to travel from the mirror back to the cavity. However, in order to optimize the composite fidelity, it is convenient to be able to introduce a time lag to the phase profile of the mirror with respect to that of the left phase shifter. We label this extra time lag for the mirror as $\delta_M$, such that $\theta(t) = \theta_1(t - \delta_M)$.
\par
Next, we relate the rightward-traveling output $R_{out}$ to the rightward-traveling input $L_{in}$ by applying the cavity input-output relationship from Eq.~\eqref{eq: left-to-right input-output} and the effect of the phase shifter on the left input as described by Eq.~\eqref{eq: left input phase shifter}:
\begin{align} \label{eq: left input to right output overall}
\begin{split}
R_{out}(t - \delta_F) &= L_{in,c}(t - \delta_F) + \sqrt{\kappa_e} A_c(t - \delta_F) \\
&= e^{-i\frac{\theta_1(t - \delta_F)}{2}} L_{in}(t - \delta_F) + \sqrt{\kappa_e} A_c(t - \delta_F)
\end{split}
\end{align}
Substituting this expression into Eq.~\eqref{eq: right output to right input}, and then substituting the result into Eq.~\eqref{eq: dot A_c setup}, we find that the differential equation for $\dot{A_c}$ reduces to a function of the cavity amplitude $A_c$ and the rightward-traveling input $L_{in}$, as desired:
\begin{align} \label{eq: dot A_c}
\begin{split}
\dot{A_c}(t) &= i \kappa_e \sin{(\theta_1(t - \delta_C))} A_c(t) - \kappa_e A_c(t) \\
&\quad - \sqrt{\kappa_e} \bigg(e^{i\theta_1(t - \delta_F/2 - \delta_M)} \Big(e^{-i\frac{\theta_1(t - \delta_F)}{2}} L_{in}(t - \delta_F) + \sqrt{\kappa_e} A_c(t - \delta_F)\Big) \bigg) - \sqrt{\kappa_e} e^{-i\frac{\theta_1(t)}{2}} L_{in}(t),
\end{split}
\end{align}
where $\delta_C$ denotes the time lag for the cavity detuning with respect to the phase shifter (i.e. $\theta_2(t) = \theta_1(t - \delta_C)$), representing another parameter which we wish to optimize. We are now ready to calculate the effect of the round trip time $\delta_F$, the mirror phase lag $\delta_M$, and the cavity detuning lag $\delta_C$ by applying our specific values for the input signal $L_{in}(t)$, phase shift $\theta(t)$, and output coupling rate for each port $\kappa_e$. As discussed in the previous section, our input signal features an exponential temporal profile:
\begin{equation}
L_{in}(t) = \sqrt{r} e^{-\frac{r}{2} t},
\end{equation}
where $r$ is the initial input rate through the left port (i.e. the inverse rise time). Substituting this into Eq.~\eqref{eq: dot A_c}, we can show that if time is converted to the dimensionless quantity $\tau' = \kappa_e t$, then the evolution of $A_c$ with respect to $\tau'$ varies only with the ratio between $r$ and $\kappa_e$:
\begin{align}
\begin{split}
\frac{dA_c}{d\tau'} &= \frac{\dot{A_c}}{\kappa_e} \\
&= i \sin{(\theta_1(\tau' - \kappa_e \delta_C))} A_c(\tau') - A_c(\tau') \\
&\quad - e^{i\theta_1(\tau' - \kappa_e \delta_F/2 - \kappa_e \delta_M)} \bigg(e^{-i\frac{\theta_1(\tau' - \kappa_e \delta_F)}{2}} \sqrt{\frac{r}{\kappa_e}} e^{-\frac{r}{2\kappa_e} (\tau' - \kappa_e \delta_F)} + A_c(\tau' - \kappa_e \delta_F) \bigg) \\
&\quad - e^{-i\frac{\theta_1(\tau')}{2}} \sqrt{\frac{r}{\kappa_e}} e^{-\frac{r}{2\kappa_e} \tau'}.
\end{split}
\end{align}
We are now in a position to numerically simulate the effect of a given mirror phase time lag $\delta_M$ and cavity detuning phase time lag $\delta_C$ by substituting a specific phase profile $\theta_1(\tau')$, a specific ratio $r/\kappa_e$ between the inverse rise time and the raw single-port cavity output coupling rate, and a given round trip time $\delta_F$ between the cavity and the mirror. Recall that the optimal phase profile is shown in Eq.~\eqref{eq: theta(tau')}, while we chose $r$ and $\kappa_e$ such that $r/\kappa_e = 1/3$. Furthermore, for our design, the minimal round trip time between the cavity and the mirror is $\delta_F = 60 \textrm{ ns}$. As such, we numerically solve for the optimal fidelity by scanning over the possible values for $\delta_M$ and $\delta_C$ and calculating $|A_c(\tau')|^2$ as $\tau' \rightarrow \infty$ for each pair of time lag parameters. Note that for $\delta_F = \delta_M = \delta_C = 0$, the simulation yields a fidelity of 96.9\%, matching the value calculated from the full-quantum method. However, for $\delta_F = 60 \textrm{ ns}$, given $\kappa_e = 2\pi \times 300 \textrm{ kHz}$, the maximum fidelity as a function of $\delta_M$, as well as the value of $\delta_C$ corresponding to this maximum, are shown in Figure 6(c) in the main text. The optimal value of the cavity detuning time lag $\delta_C$ relates linearly with the mirror time lag $\delta_M$. For the given range, we find that $\delta_M \ge 0$ and $\delta_C \le 0$, indicating that while the mirror clock lags the left phase shifter clock, the cavity detuning clock actually \textit{leads} the left phase shifter clock. It is evident that the fidelity is maximized if $\delta_M = 21 \textrm{ ns}$ and $\delta_C = -34 \textrm{ ns}$, at a value of approximately 89.0\%. 

\section{Tunable-reflectivity mirror}

We can point to one final application of the phase shifter device platform. Since the guided mode has increasing dispersion across the entire Brillouin zone, operating points near the center and edge of the Brillouin zone (at $k_x=0$ or $k_x = \pi/a$) can be used to construct mirrors with tunable reflectivity because the frequency of the signal can be shifted into the gap of unallowed frequencies. The principle of operation is illustrated in \figureautorefname{ \ref{fig:mirror_bands}}(a). When the frequency of the operating point is shifted off the band entirely, the signal cannot propagate in the waveguide and will be reflected. To implement this at the same operating frequency as the other devices previously described, we can adiabatically increase the minor axis of the ellipse at the center of the waveguide to $1.06 \times 0.4h$. For this new geometry, the propagation constant at the operating frequency shifts to approximately $0.97\pi/a$ and the group velocity falls to 34 m/s. The variation of $\Delta f$ with piezo bias is shown in \figureautorefname{ \ref{fig:mirror_bands}}(b). Since $\Delta f$ exceeds the MHz range at moderate voltages and the band edge at $k_x = \pi/a$ is only 486 kHz greater than the operating frequency, the tunable mirror can be achieved at moderate bias ($\Delta f = -685$ kHz at $-5$ V). In addition, the bandstructure of the waveguide retains its desirable isolation in frequency, even for the perturbed geometry [\figureautorefname{ \ref{fig:mirror_bands}}(c)].
\begin{figure}
	\centering
	\includegraphics[width=\textwidth]{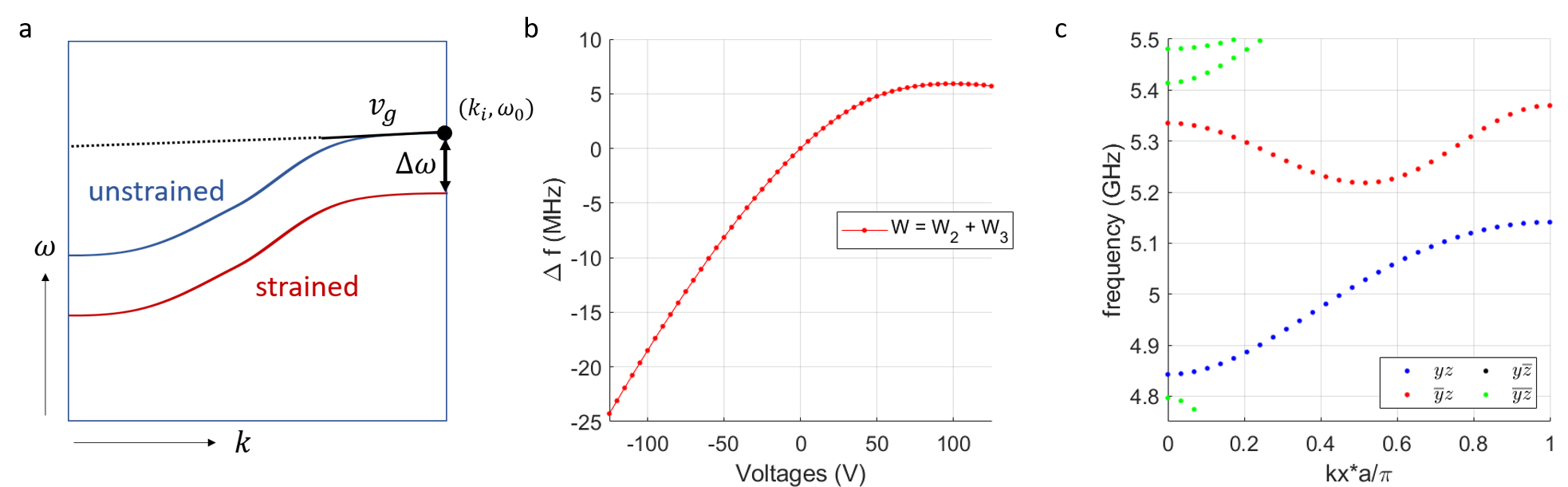}
	\caption{(a) Operating principle for a mirror with tunable reflectivity. (b) $\Delta f$ versus piezo bias for a waveguide with a 6\% expansion in the minor axis of the central ellipse. Here, $k_x = 0.97\pi/a$. (c) Bandstructure computed for the new geometry. With the minor structural modification, the guided mode retains its symmetry and isolation in frequency.}
	\label{fig:mirror_bands}
\end{figure}

\section{Tailoring the waveguide's operating point}

In the context of the the tunable mirror, we noted that it is possible to tailor the structure of the waveguide to optimize the dispersion at the operating point and its position within the Brillouin zone. In particular, when the minor axis of the ellipse is increased by a few percent, the figure of merit of the phase shifter drops. In some cases, this reduced phase shifter operating voltage or device length may be advantageous. The process for finding the new propagation constant is shown in \figureautorefname{ \ref{fig:reduced_vg_op_pt}}. The bandstructure was solved by discretizing $k_x$ into 30 equally-spaced points across the Brillouin zone. Then the frequencies were fitted with a cubic spline function which can be used to identify interpolated values. At the 5.1406 GHz frequency of the band, the propagation constant $k_x=0.8704\pi/a$. This operating point is shown in \figureautorefname{ \ref{fig:reduced_vg_op_pt}}(a) by the large, blue, closed circle. The group velocity at the operating point is $v_g = 138$ m/s, which is plotted in red. \figureautorefname{ \ref{fig:reduced_vg_op_pt}}(b) shows the recomputed trend in the length required to achieve a $\pm\pi$ phase shift. Since the geometrical modification does not significantly change the magnitude of the frequency shift $\Delta f$, the difference in the length required for the phase shift arises principally due to the reduced group velocity. With this new geometry, the voltage required for a 100 period phase shifter to access $\pm\pi$ phase shifts drops to the magnitude of 10 volts, as shown in \figureautorefname{ \ref{fig:reduced_vg_op_pt}}(c). Since the only difference in the waveguide geometry is a small perturbation, adiabatic structural modifiactions between these waveguide structures for tuning the dispersion between components in a larger network of devices is straightforward.
\begin{figure}
	\centering
	\includegraphics[width=\textwidth]{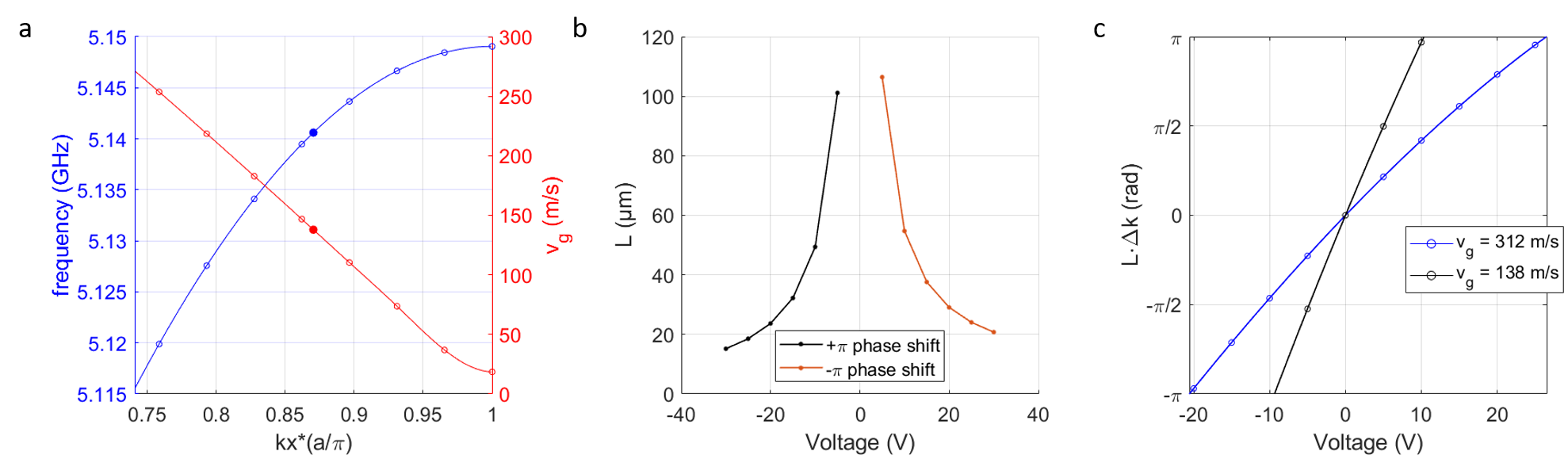}
	\caption{(a) Numerical identification of the new propagation constant obtained when the minor axis of the defect ellipse increases by 5\%. The new operating point at the frequency of 5.1406 GHz is shown by the filled circles. (b) Results from simulating the phase shifting response at the new operating point. The length of strained waveguide required to obtain a $\pm\pi$ frequency shift at a given voltage has been reduced by over a factor of 2, consistent with the reduction in group velocity. (c) Comparison of the control biases needed for a phase shifting waveguide with reduced group velocity compared to the one presented in the main text.}
	\label{fig:reduced_vg_op_pt}
\end{figure}

\bibliography{Piezo-acoustomechanical.bib}

\end{document}